# A Short Introduction to Plasma Physics

*K. Wiesemann*
AEPT, Ruhr-Universität Bochum, Germany

**Abstract**
This chapter contains a short discussion of some fundamental plasma phenomena. In section 2 we introduce key plasma properties like quasi-neutrality, shielding, particle transport processes and sheath formation. In section 3 we describe the simplest plasma models: collective phenomena (drifts) deduced from single-particle trajectories and fundamentals of plasma fluid dynamics. The last section discusses wave phenomena in homogeneous, unbounded, cold plasma.

## 1 Introduction

Plasma exists in many forms in nature and has a widespread use in science and technology. It is a special kind of ionized gas and in general consists of:

– positively charged ions ('positive ions'),

– electrons, and

– neutrals (atoms, molecules, radicals).

(Under special conditions, plasma may also contain negative ions. But here we will not discuss this case further. Thus in what follows the term 'ion' always means 'positive ion'.) We call an ionized gas 'plasma' if it is *quasi-neutral* and its properties are dominated by electric and/or magnetic forces.

Owing to the presence of free ions, using plasma for ion sources is quite natural. For this special case, plasma is produced by a suitable form of low-pressure gas discharge. The resulting plasma is usually characterized as 'cold plasma', though the electrons may have temperatures of several tens of thousands of Kelvins (i.e. much hotter than the surface of the Sun), while ions and the neutral gas are more or less warm. However, owing to their extremely low mass, electrons cannot transfer much of their thermal energy as heat to the heavier plasma components or to the enclosing walls. Thus this type of cold plasma does not transfer much heat to its environment and it may be more exactly characterized as 'low-enthalpy plasma'.

## 2 Key plasma properties

### 2.1 Particle densities

Owing to the presence of free charge carriers, plasma reacts to electromagnetic fields, conducts electrical current, and possesses a well-defined space potential.

Positive ions may be singly charged or multiply charged. For a plasma containing only singly charged ions, the ion population is adequately described by the ion density $n_i$,

$$n_i = \frac{\text{number of particles}}{\text{volume}}, \quad [n_i] = \text{cm}^{-3} \text{ or } [n_i] = \text{m}^{-3}. \qquad (1)$$

Besides the ion density, we characterize a plasma by its electron density $n_e$ and the neutral density $n_a$.

## 2.2 Ionization degree, quasi-neutrality

Quasi-neutrality of a plasma means that the densities of negative and positive charges are (almost) equal. In the case of plasma containing only singly charged ions, this means that

$$n_i \approx n_e. \qquad (2)$$

In the presence of multiply charged ions, we have to modify this relation. If $z$ is the charge number of a positive ion and $n_z$ is the density of $z$-times charged ions, the condition of quasi-neutrality reads

$$n_e \approx \sum_z z \cdot n_z. \qquad (3)$$

The degree of ionization is defined with the particle densities, not with the charge densities. However, there are two different definitions in use:

$$\eta_i = \frac{\sum_z n_z}{n_a + \sum_z n_z} \quad \text{and} \quad \eta_i' = \frac{\sum_z n_z}{n_a}. \qquad (4)$$

Strictly speaking, $\eta_i'$ is an approximation of $\eta_i$ for $\eta_i \ll 1$, which is the usual case. Typical values of $\eta_i$ for plasmas in ion sources are in the range of $10^{-5}$ to $10^{-3}$. Fully ionized plasma corresponds to $\eta_i = 1$ ($\eta_i' \to \infty$ in this case).

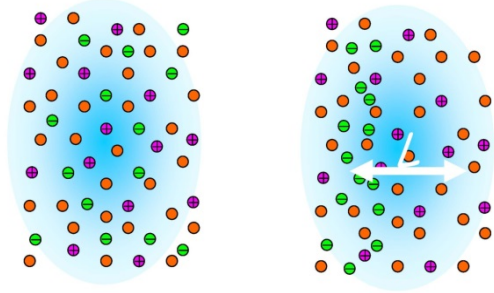

**Fig. 1:** Charge separation, schematic

To investigate quasi-neutrality further, we assume that a cloud of electrons in plasma has moved to a certain area, forming a negative space charge there. A similar ion cloud is left without electrons in a distance $L \approx \delta x$, forming a positive space charge (see Fig. 1). Thus one obtains between these space-charge clouds an electric field having its maximum value $\overrightarrow{E_{max}}$ at the mutual borders. We can estimate the value of $\overrightarrow{E_{max}}$ by using Poisson's equation:

$$E_{max} = \frac{e \cdot n_i \cdot \delta x}{\varepsilon_0}. \qquad (5)$$

The direction of $E$ is such that the electric force drives the two clouds back to overlap. Here $e$ is the positive elementary charge and $\varepsilon_0$ is the permittivity of free space.

For further discussion, we calculate the gain in potential energy $W_{pot}$ of a charged particle after moving by $\delta x$ through the space-charge layer:

$$W_{pot} = \int_0^{\delta x} eE\, dx = \frac{e^2 n_e (\delta x)^2}{2\varepsilon_0}. \qquad (6)$$

The only energy available for this purpose is the thermal energy of electrons (and ions, but, owing to their usually low temperature compared to electrons, ion thermal effects can be neglected in cold plasma, i.e. in the plasma of interest for ion sources), at the average $\tfrac{1}{2}k_B T_e$ for a movement in one degree of freedom. Thus we may expect deviations from quasi-neutrality on a scale defined by

$$W_{pot} = \frac{1}{2} k_B T_e. \qquad (7)$$

This corresponds to a charge separation over the so-called Debye–Hückel length $\lambda_D$ [1] given by

$$\lambda_D = \left( \frac{\varepsilon_0 k_B T_e}{e^2 n_e} \right)^{1/2}. \qquad (8)$$

A numerical value of this length is given by $\lambda_D / \mathrm{m} = 7.434 \times 10^3 (k_B T_e / \mathrm{eV})^{1/2} / (n_e / \mathrm{m}^{-3})^{1/2}$ (see also Fig. 2).

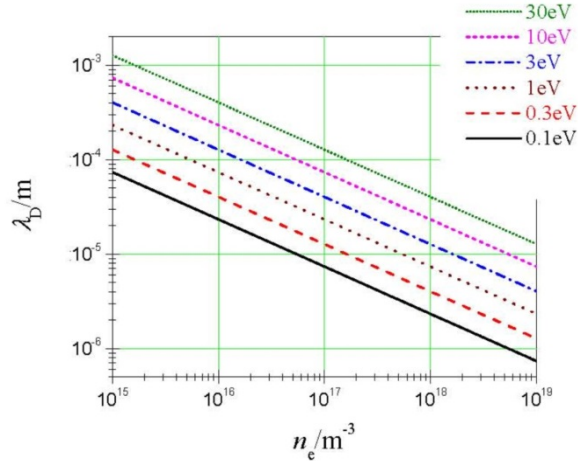

**Fig. 2:** Debye length versus plasma density and electron temperature

We may, on the other hand, ask what amount $\Delta n = |n_e - n_i|$ of deviation from quasi-neutrality is possible over a given length $L$. Again, we have only the thermal energy at our disposal. Thus

$$\frac{1}{2} k_B T_e \approx \frac{1}{2} \frac{e^2}{\varepsilon_0} \cdot \Delta n \cdot L. \qquad (9)$$

When substituting $k_B T_e$ by $\lambda_D$ we obtain as an estimate

$$\frac{\Delta n}{n} \approx \left(\frac{\lambda_D}{L}\right)^2. \tag{10}$$

We may formulate the condition of quasi-neutrality as $\Delta n \ll n_e, n_i$. According to Eq. (10), this is equivalent to $L \gg \lambda_D$. This means that the extension of an ionized gas must be large compared to the Debye–Hückel length, in order to fulfil the conditions of being plasma. Plasma quasi-neutrality is defined only on a large macroscopic scale. If we inspect plasma on a microscopic scale, we may find deviations from neutrality increasing with decreasing scale length.

In plasmas powered electrically (i.e. in any electric discharge), the electrons gain energy more easily from the external electric field than the inert ions, as the electrons are much lighter and thus electronic currents are much larger. Since in elastic collisions electrons can transfer kinetic energy only in small amounts – of the order of $m_e/m_i$ – to the ions ($m_e$ and $m_i$ are the electronic and ionic masses, respectively), in steady state the electron temperature will be much higher than the ion (and neutral) temperatures, as discussed above. Thus, electrons are mainly responsible for local deviations from neutrality and the temperature in the formula of the Debye–Hückel length is the electron temperature $T_e$.

In a typical low-power ion source plasma, the electron temperature $T_e$ is of the order 30 000– 40 000 K, while the ion temperature $T_i$ is around 500–1000 K. The electron density $n_e$ amounts to about $10^{10} \text{cm}^{-3} = 10^{16} \text{m}^{-3}$ and higher. Under these conditions, the Debye–Hückel length is of the order of 0.12–0.16 mm and shorter.

(In many plasma physics texts, the symbol $T$ stands for the product of temperature with the Boltzmann constant $k_B$, i.e. for the energy $k_B T$ measured in electronvolts (eV), instead of for the thermodynamic temperature. This characteristic energy is dubbed 'temperature measured in eV'. A temperature of 11 600 K corresponds to a characteristic energy of 1 eV.)

### 2.3 Plasma oscillations

The value of the electric field $\vec{E}$ created by charge separation is, as we have seen, proportional to the separation length, which we now call $x$:

$$E = \frac{e}{\varepsilon_0} nx. \tag{11}$$

Thus we obtain for the movement of, say, electrons, under the action of the restoring force $F = eE$,

$$F = eE = \frac{e^2}{\varepsilon_0} nx = m_e \frac{d^2 x}{dt^2}. \tag{12}$$

This is the equation of a harmonic oscillator with the eigenfrequency

$$\omega_{pe} = \left(\frac{e^2 n}{\varepsilon_0 m_e}\right)^{1/2}, \tag{13}$$

the so-called (angular) electron plasma frequency. A numerical value of the (electron) plasma frequency is given by $\omega_{pe} = 2\pi \times 8.9 \times \sqrt{n_e / \text{m}^{-3}}$. For the plasma data given above, this yields $\omega_{pe} = 2\pi \times 8.9 \times 10^8 \text{s}^{-1}$.

What we describe here are oscillations of the electron charge cloud as a whole (see Figs. 1 and 3). The inert ions are considered to remain at rest. A more careful analysis will reveal that, instead of these oscillations, different types of acoustic waves can propagate in plasma. However, the electron and ion plasma frequencies will show up as important parameters for characterizing these different types of plasma waves.

By replacing the electron mass by the ion mass in Eq. (13), one obtains the (angular) ion plasma frequency. It is the natural frequency of ion space charge and may play a role in the ion sheaths in front of a wall or between plasma meniscus and extraction hole at the output of an ion source. In plasma, ion acoustic waves are strongly damped at this frequency – see the discussion below.

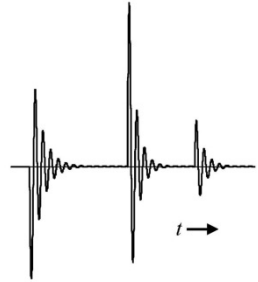

**Fig. 3:** Plasma oscillations

Our considerations show that neutrality is a dynamic equilibrium state of plasma from which deviations are possible on time scales defined by the (electron) plasma frequency and extending over spatial dimensions of the order of the Debye–Hückel length. These deviations are powered by the thermal energy of the charged plasma constituents and tend to decay into the neutral equilibrium state. Thus plasmas are always close to neutrality – they are quasi-neutral.

## 2.4 Plasma as a gas

A gas is described adequately by single-particle properties averaged over the particle distribution functions and parameters like pressure, temperature and density, which can be correlated to those averages, as we know from kinetic theory.

Plasma kinetic theory is classical Boltzmann statistics, if the distance between particles (electrons, ions, neutrals) is sufficiently large (*classical plasma*). For electrons, this is the case if their average distance,

$$\lambda_n = 1/(n_e)^{1/3}, \tag{14}$$

is large compared to the average electron de Broglie wavelength $\lambda_B$,

$$\lambda_B = h/m_e v_{th}, \text{ with } \tfrac{1}{2}m_e v_{th}^2 = k_B T_e. \tag{15}$$

Otherwise plasma is *degenerate*.

A plasma can be described as an *ideal gas* if the mutual potential energy of electrons and ions is small compared to the average kinetic energy $\tfrac{3}{2}k_B T_e$, that is, if

$$\frac{3}{2}k_\text{B}T_\text{e} \gg \frac{e^2}{4\pi\varepsilon_0 \lambda_\text{n}}.  \qquad (16)$$

Substituting $k_\text{B}T_\text{e}$ by the Debye–Hückel length, we obtain as equivalent conditions:

$$\lambda_\text{D} \gg \lambda_\text{n} = \frac{1}{n_\text{e}^{1/3}} \qquad (17)$$

or

$$g = \left(\frac{1}{n_\text{e}}\frac{4\pi}{3}\lambda_D^3\right)^{-1} \ll 1. \qquad (18)$$

The expression in the bracket is the number of electrons in a so-called Debye–Hückel sphere, that is, in a sphere with the volume $\frac{4\pi}{3}\lambda_\text{D}^3$. Its reciprocal $g$ is the 'plasma parameter', taken in plasma theory as a measure for degeneracy or absence of degeneracy in plasma.

Plasma with large plasma parameter is 'non-ideal' or 'strongly coupled classical plasma'. Now

$$g \propto \frac{n_\text{e}^{1/2}}{(k_\text{B}T_\text{e})^{3/2}}. \qquad (19)$$

Thus non-ideal classical plasma is very cold and very dense. In such a case, correlations between the plasma particles may become important. Under laboratory conditions, such correlations can be observed in dusty plasmas, where dust particles sometimes adjust themselves into regular structures. In ion source plasma we have $g \ll 1$ as a rule. Such plasma behaves classically, that is, obeys classical Boltzmann statistics, it is in general non-degenerate and the conditions in Eqs. (17) and (18) are fulfilled.

### 2.5 Particle transport in plasma

We restrict our discussion to drift and diffusion, transport processes important in ion source plasma. These processes are characterized by 'transport coefficients', of which we discuss mobility $b$, conductivity $\sigma$ and diffusion coefficient $D$.

#### 2.5.1 Mobility and conductivity

To understand the concept of mobility, we consider a simple method for measuring viscosity in a viscous fluid, the falling sphere method. Under the action of gravity, a small metallic sphere will, after a short distance, fall with constant speed, which can be taken as a measure of viscosity. In ion source plasma, viscosity is a higher-order effect and can be neglected. We will not consider it further. However, under the common action of a constant force due to, say, an electric field $\vec{E}$ and friction due to collisions with other particles, a charged particle may attain a constant drift velocity $\vec{v}_\text{D}$, which under favourable conditions is proportional to the value of the force acting, that is, to $qE$. Here $q$ is the particle charge. The proportionality constant is the *mobility* $b$:

$$\vec{v}_\text{D} = bq\vec{E}. \qquad (20)$$

Here we follow the mobility definition given by Allis [2]. However, a warning: many authors use a different mobility definition:

$$\vec{v}_D = \tilde{b}\vec{E}. \tag{21}$$

The advantage of the definition according to Eq. (20) is that $b$ is positive by definition, whereas $\tilde{b}$ carries the sign of the charge $q$, that is, $\tilde{b}$ is negative for electrons and positive for positive ions, thus causing some complications, especially when defining the conductivity.

The density of the electric current $\vec{j}_e$ carried by drifting charged particles is given by

$$\vec{j}_e = qn_q\vec{v}_D = q^2n_q\vec{E} = \sigma\vec{E}. \tag{22}$$

Here we have used Eq. (20) to eliminate $\vec{v}_D$ and to define the *conductivity* $\sigma$. In our definition the conductivity also comes out to be positive by definition.

In weakly ionized plasma, friction is due to charged particle collisions with neutrals. For electrons, one obtains from kinetic theory

$$b_e = \langle 1/m_e\nu_{en}\rangle = \langle 1/m_e Q_{en} v_e\rangle \propto 1/\sqrt{m_e}. \tag{23}$$

Here the angle brackets indicate an average over the electron distribution function, $\nu_{en}$ is the collision frequency for momentum transfer collisions between electrons and neutrals, $Q_{en}$ is the respective cross-section for momentum transfer, and $v_e$ is the velocity of a single electron. The averages are defined by

$$\langle 1/m_e Q_{en} v_{en}\rangle = \iiint_{v_e} \left(1/m_e Q_{en} v_{en}\right) f(\vec{v}_e)\, d^3 v_e, \tag{24}$$

where $f(\vec{v}_e)$ is the electron velocity distribution function. Our formulas are valid if the value of drift velocity $\vec{v}_D$ is small compared to the average value of the electron velocity

$$\langle v_e\rangle = \sqrt{8m_e/\pi k_B T_e}. \tag{25}$$

Otherwise $f(\vec{v}_e)$ must be replaced by the distribution function of the drifting electrons, and $b$ and $\sigma$ will become functions of the electron drift velocity $v_{eD}$. Equation (20) describes a stationary equilibrium between electric force and friction. Under the condition of this equilibrium, the energy taken by the drifting particles is completely transformed into heat of those particles. In fusion research, this process constitutes an important mechanism for plasma (electron) heating and is called 'ohmic heating'. In the case of fully ionized fusion plasma, the necessary friction is produced by electron–ion collisions.

A detailed analysis reveals that such equilibrium is possible only if the decrease of the friction force with increasing absolute value of $v_{eD}$ is sufficiently slow. The friction force decreases because of the decrease of the Coulomb collision cross-sections with increasing collision energy. In fully ionized fusion plasma, the condition for equilibrium is fulfilled for electrons as long as

$$\tfrac{1}{2} m_e v_{eD}^2 \leq k_b T_e . \tag{26}$$

Otherwise electrons may be continuously accelerated to relativistic energies, an effect known as 'run-away' [3]. To avoid run-away, the drift velocity must be guided to rise sufficiently slowly that Eq. (26) is never violated. A similar effect may also occur in weakly ionized plasma because all collision cross-sections with neutrals decrease at higher energies [4]. In general, the equilibrium between collisional friction and an external force is a phenomenon at low energies.

### 2.5.2 Diffusion

Let us consider a virtual plane somewhere in homogeneous plasma. Owing to Brownian motion, there is a continuous flow of electrons, ions and neutrals from both sides through this plane. According to a famous formula of gas kinetics, its (particle) current density $\Gamma$ is given by

$$\Gamma = n_q \sqrt{m_e / 2\pi k_B T_e} . \tag{27}$$

As the particle densities and temperatures in homogeneous plasma are equal on both sides, these flows will also be equal for any particle sort and compensate each other.

In the case of density (and temperature) gradients, however, the flows no longer equal each other, and a net flow results in the direction of decreasing particle density. This transport process is *diffusion*. It can be described by the equations known as Fick's laws. According to the first of Fick's laws, the diffusive particle flux $\overrightarrow{\Gamma_{diff}}$ is given by

$$\overrightarrow{\Gamma_{diff}} = -D\nabla n . \tag{28}$$

It follows from our discussion that the value of the current density of the diffusive flux is always smaller than the flux given in Eq. (27).

The diffusion constant $D$ is related to the respective mobility by the famous Einstein relation

$$D_{e,i} = b_{e,i} k_B T_{e,i} = k_B T_{ei} / m_{ren,in} \nu_{en,in} \approx \langle v_{e,i}^2 \rangle / \nu_{en,in} \approx \nu_{en,in} \langle \lambda_{en,in} \rangle^2 , \tag{29}$$

if the mobility definition of Eq. (20) is used. Here the brackets $\langle \cdots \rangle$ stand for an average over the particle distribution function. The last two approximate expressions in Eq. (29) are very useful for estimating orders of magnitude. From relations (29) and Eq. (23) and the respective formula for the ion mobility it follows that

$$D_e / D_i = \sqrt{m_{rin} / m_e} , \tag{30}$$

that is, in weakly ionized (and non-magnetized) plasma, electrons diffuse much faster than ions. Here $m_{rin}$ is the reduced mass of the ion–neutral collision system as used in collision theory. Also $m_e$ stands for the reduced mass of the electron–neutral collision system here and in Eq. (23). However $m_{re} \approx m_e$ because $m_e$ is so small. Equation (29) is in principle also valid in fully ionized plasma, if that is sufficiently close to equilibrium. In that case, the collision frequencies of electrons and ions with neutrals must be replaced by the electron–ion collision frequencies. 'Close to equilibrium' means that the distribution function of the diffusing particles is very close to a Maxwellian and any drift velocity is very small. In more complicated situations, $D$ must be calculated from kinetic theory, which is not treated here. The interested reader may find a discussion in books on plasma kinetic theory – see for instance Ref. [5].

## 2.6 Sheath formation

Ion source plasma is enclosed by a vessel. Its walls are sinks for charged particles, causing continuous flows of electrons and ions to the walls.

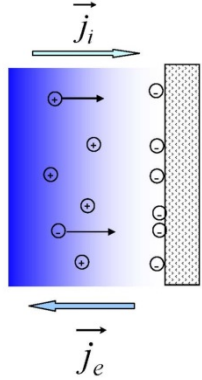

**Fig. 4:** Electron and ion flows from plasma to a wall (schematic)

Ions may stick on or in a wall or be neutralized and re-emitted as neutrals. Electrons may stick on a dielectric wall and bind in low-energy surface states, be absorbed by metallic walls, or be re-emitted and reflected. Further ion (and metastable) and electron bombardment of a wall may cause secondary electron emission. However, in our simplified discussion, we will neglect back-flows from the walls as a secondary effect. The electron and ion flows towards a wall correspond to electric currents with antiparallel current densities (see Fig. 4). Owing to their low mass, the transport of electrons is much faster than ion transport. This leads in the steady state to the formation of a tiny positive space charge in plasma and a negative charge on a wall. Because of these charges, the plasma potential is positive and the wall potential is negative. In the case of a dielectric wall, the electric current to the wall must be zero, the same holding true for the electric current density $\vec{j}$:

$$\vec{j} = \vec{j_e} + \vec{j_i} = 0. \tag{31}$$

The charges in plasma and on the walls regulate themselves in such a way that Eq. (31) is fulfilled. Assuming the plasma potential to be zero and that the electrons obey a Maxwell distribution function, we can rewrite Eq. (31) as

$$\vec{j_i} = -\vec{j_e} = -en_e \left(\frac{m_e}{2\pi k_B T_e}\right) \exp\left(\frac{eU_{wall}}{k_B T_e}\right). \tag{32}$$

Here $e$ is the *positive* elementary charge. Using Eq. (32) we may easily calculate the wall potential $U_{wall}$. (It is also equal to the potential of a floating probe.) In the case of metallic walls, Eq. (31) may not necessarily be fulfilled, only the respective condition for the total electric currents. This is of special importance for the case of magnetized plasma. In this case diffusion across the magnetic field differs from diffusion along the magnetic field lines. Thus compensating currents may flow in a metallic vessel. These currents greatly influence plasma containment in such cases. We will discuss this problem further in section 3.2.

# 3 Plasma modelling

## 3.1 Movement of charged particles under the action of electric and magnetic fields

Consider a single singly charged particle (mass $m$, charge $q$) in a combination of a stationary force field $\vec{F}$ not depending on the particle velocity (like the force $q\vec{E}$ in an electric field $\vec{E}$), and a stationary magnetic field with induction $\vec{B}$. We denote by $\vec{v}$ the velocity of this particle and by $\dot{\vec{v}}$ the time derivative of velocity, its acceleration. The equation of motion in vector notation,

$$m\dot{\vec{v}} = q\vec{v} \times \vec{B} + \vec{F}, \tag{33}$$

constitutes a set of three differential equations. The first term on the right-hand side has only components perpendicular to $\vec{B}$. Thus the equation for the motion in the direction parallel to $\vec{B}$ is independent of $\vec{B}$ and describes a constant acceleration parallel to $\vec{B}$ by the respective component of $\vec{F}$, denoted by $\vec{F}_\parallel$. The components of $\vec{F}$ and $\vec{v}$ perpendicular to $\vec{B}$ and their time derivatives we will term $\vec{F}_\perp$, $\vec{v}_\perp$, and so on.

Thus, *perpendicular* to $\vec{B}$, we have a system of two coupled inhomogeneous differential equations:

$$m\dot{\vec{v}}_\perp = q\vec{v}_\perp \times \vec{B} + \vec{F}_\perp. \tag{34}$$

The solution is a combination of the general solution of the homogeneous part of these equations and a special solution of the inhomogeneous equation. The homogeneous part is given by

$$m\dot{\vec{v}}_\perp = q\vec{v}_\perp \times \vec{B}. \tag{35}$$

It describes the motion of our particle under the action of the magnetic field only. The solution is well known. The motion is a gyration in the plane perpendicular to $\vec{B}$, that is, a motion with constant velocity on a circle, the so-called cyclotron motion (see Fig. 5). The radius of the circle, the cyclotron radius, $r_B$, is given by

$$r_B = \frac{mv_\perp}{qB}. \tag{36}$$

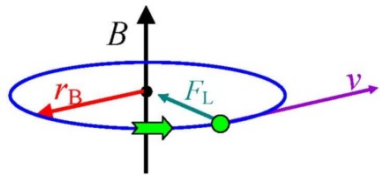

**Fig. 5:** Cyclotron motion of a charged particle

Here $v_\perp$ is the speed of the particle in the plane perpendicular to $\vec{B}$, thus $mv_\perp = p_\perp$ is the absolute value of the respective momentum. The respective angular frequency, the cyclotron frequency $\omega_B$ is given by

$$\omega_B = \frac{v_\perp}{r_B} = \frac{qB}{m}. \tag{37}$$

It does not depend on the particle velocity – at least in the non-relativistic case. Note that the charge $q$ may be either positive or negative. In our definition, $\omega_B$ has the same sign – it defines the rotational direction.

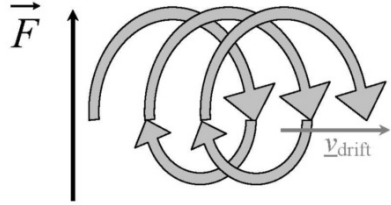

**Fig. 6:** Drift of a charged particle due to the combined action of a constant magnetic field with the induction $B$ and a constant force $F$.

We obtain a special solution of the inhomogeneous equation by assuming the velocity $\vec{v}_\perp$ to be constant, which means zero acceleration. In this case we obtain

$$-q\vec{v}_\perp \times \vec{B} = \vec{F}_\perp . \tag{38}$$

To solve for the velocity we generate the cross-product of this equation with $\vec{B}$ from the right, yielding

$$qB^2 \vec{v}_\perp = \vec{F}_\perp \times \vec{B} \quad \text{or} \quad \vec{v}_\perp = \frac{\vec{F} \times \vec{B}}{qB^2} \equiv \vec{v}_{\text{drift}}. \tag{39}$$

The complete particle motion in a plane perpendicular to $\vec{B}$ is a superposition of a gyration and a constant velocity $\vec{v}_{\text{drift}}$ in the direction perpendicular to $\vec{B}$ and $\vec{F}$. The cyclotron motion depends on the initial velocities of the particles. These velocities are distributed at random. When we average over all particles, the mean value of the gyration velocity will be zero. In contrast to that, at least for all particles of equal charge $\vec{v}_{\text{drift}}$ is the same. Its value and direction depend only on the direction and magnitude of the magnetic induction and the external force $\vec{F}$. The whole population of equal particles will drift in the same direction. In this way we have reduced the discussion of particle motion to the discussion of the motion of the guiding centre for gyration as a pseudo-particle. This strategy is known as the guiding centre approximation. Most important is the case of the force due to an electric field, $\vec{F} = q\vec{E}$. In this case the charge cancels out in Eq. (39). All charged particles drift with the same velocity in the same direction ($E \times B$ drift).

We have pictured this drift by considering the influence of the external force on the gyration (see Fig. 6): imagine a magnetic field with the field lines pointing perpendicular into the plane of the paper. Under this condition, the gyration is restricted to the plane of the paper (or a plane parallel to it). The case shown in Fig. 6 is that of a particle with positive charge under the influence of an electric field. When the particle moves upwards, it is accelerated and the radius of its gyration circle increases. Downwards it is decelerated and the radius of the gyration circle decreases. Thus its trajectory is a cycloid instead of a closed circle.

In the case of a negatively charged particle, the direction of the gyration is reversed. If the force is due to an electric field, the direction of the force $F$ is also reversed and the drift, therefore, goes in the same direction as before. If the direction of the force is independent of the particle charge (like, for example, in the case of inertial or gravitational forces), the direction of the drift will be reversed. In this latter case positively and negatively charged particles drift in opposite directions. The result may be an electric current or charge separation.

If only the magnetic field strength decreases in the direction indicated by the arrow for *F*, we get a similar effect because of the dependence of the gyration radius on *B*. This so-called gradient drift is also an example of current transport by a drift. Any effect changing the gyration radii in a similar way can thus cause a drift. A different approach to drift induced by a gradient of the magnetic induction *B* is by considering the gyrating particle as a pseudo-particle with a magnetic moment $\vec{M}$. In analogy to a current-carrying wire loop, we can treat the gyrating particle as a circular current $I = q/\tau_B = q\omega_B/2\pi$. The area *A* within the loop is $A = r_B^2 \pi$. Then

$$M = I \cdot A. \tag{40}$$

This is a magnetic dipole experiencing a force in the presence of a magnetic field gradient. This force can be considered as the reason for the drift and can be treated in a similar way as we have treated the force due to an electric field. The direction of $\vec{M}$ is always opposite to the direction of $\vec{B}$ (see Fig. 7). The plasma is *diamagnetic*!

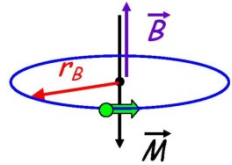

**Fig. 7:** Magnetic moment of a gyrating charged particle

If the direction of the magnetic field gradient is parallel to $\vec{B}$, we have a special effect. It turns out that the magnetic moment due to a gyrating particle moving along $\vec{B}$ is constant if the gradient of $\vec{B}$ is sufficiently small (i.e. $\vec{M}$ is an *adiabatic invariant*). Using the relations for the current *I* and the area of the gyration circle *A* given above, we obtain

$$M = I \cdot A = W_\perp / B. \tag{41}$$

Here $W_\perp = \tfrac{1}{2} m v_\perp^2$ is the kinetic energy due to the gyration of the considered charged particle, that is, due to the motion perpendicular to the direction of $\vec{B}$.

Further (assuming the potential energy to be zero), the total energy

$$W = W_\perp + W_\parallel \tag{42}$$

is also a constant of motion. If the particle moves in the direction of increasing magnetic field, $W_\perp$ must increase, because *M* is constant. Thus, $W_\parallel$, the energy of the motion parallel to $\vec{B}$, must decrease by the same amount to keep *W* constant. If $W_\parallel$ becomes zero, the particle cannot proceed further and must return. A configuration with *B* increasing along *B* is therefore called a magnetic mirror (Fig. 8).

In our discussion we have tacitly assumed that collisions between plasma particles are not important. If collisions are very frequent, cyclotron motion and drift will be disturbed. We can estimate the condition for drift by comparing the average frequency of gyration, $\omega_B = v_{th}/r_B = qB/m$, with the average collision frequency $\nu_c$ (note the difference in symbols: frequency $\nu$ and velocity $v$). If at least for one kind of charged particle $\omega_B \gg \nu_c$, drift and gyration are fully developed for those particles. We call such plasma *magnetized*. Magnetization can, at least in principle, always be attained by a sufficiently strong external magnetic field. For a given magnetic field, magnetization depends on plasma and neutral density. It is strongest in dilute plasma at low pressure.

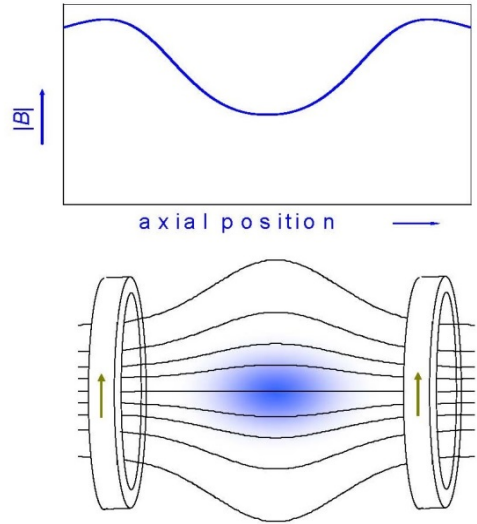

**Fig. 8:** Magnetic mirror configuration showing the shape of the magnetic field lines on a plane through the axis of the system (below) and the curve of the absolute value of the induction $\vec{B}$ on the axis of the system.

### 3.2 Diffusion in magnetized plasma

Under the action of an external magnetic field with induction $\vec{B}$, plasma becomes anisotropic. One consequence is that there is a difference between transport, such as diffusion, along and across the magnetic field. Along the magnetic field, diffusion resembles transport in non-magnetized plasma. Let $D_{\|e}$ and $D_{\|i}$ be the electron and ion diffusion coefficients for this case. They obey the same relation as in the absence of a magnetic field (see Eq. (30)):

$$\frac{D_{\|e}}{D_{\|i}} = \sqrt{\frac{m_{ri}}{m_e}}, \qquad (43)$$

the same holding true for mobility and conductivity.

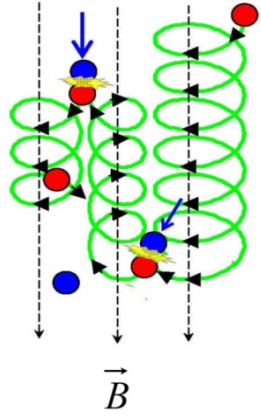

**Fig. 9:** Trajectories of the Brownian motion of a charged particle in a magnetic field

To understand diffusion across the magnetic field, we first consider the Brownian motion of, say, electrons in magnetized plasma (see Fig. 9). In Brownian motion, the trajectories of neutral particles will be straight between collisions and are bent by collisions. The resulting trajectory is a random zigzag path. In plasma, a strict distinction between free motion and collision is not possible: the trajectories of charged particles may be curved due the far-reaching Coulomb field.

In the presence of an external magnetic field (induction $\vec{B}$), free-moving charged particles spiral around a magnetic field line. If the cyclotron frequency of the spiralling particles is large compared to any collision frequency, one has the situation sketched in Fig. 9. Owing to collisions, particles hop from field line to field line. As discussed above, the picture of separate collisions and free flight in between must be modified in plasma. Here the particle motion is an $E \times B$ drift under the action of a randomly fluctuating electric field. However, calculation of diffusion coefficients leads to similar results in both models.

Taking the situation in Fig. 9 as given, we can conclude that the average distance travelled between successive collisions across the magnetic field is not the mean free path but the average cyclotron radius $\langle r_B \rangle$ – compare Eq. (36). Thus we may estimate the diffusion coefficient for diffusion across the magnetic field by replacing in the last expression $\langle \lambda \rangle$ on the right-hand side of Eq. (29) by $\langle r_B \rangle$:

$$D_{\perp e,i} \approx \nu \langle r_{Be,i} \rangle^2 \propto \sqrt{m_{e,i}} \,. \tag{44}$$

Note that the masses in Eq. (44) are *not* the reduced masses! For the ratio of the diffusion coefficients, instead of Eq. (43) we obtain

$$\frac{D_{\perp e}}{D_{\perp i}} = \sqrt{\frac{m_e}{m_i}} \,. \tag{45}$$

Across the magnetic field, ion transport is much faster than electron transport. For plasma confined magnetically in a closed vessel, this has serious consequences. If the vessel has dielectric walls, regions hit by magnetic field lines will charge up negatively, other regions positively. The charges build up a potential, repelling the fast component so much that both kinds of particle hit the wall at the same rate. Thus transport (i.e. plasma losses) is ruled by the slowly transported species:

along the field lines this is the ions, and across the field lines it is the electrons. In the case of a metallic wall, compensating currents between the different wall regions inhibit the build-up of surface charges. Thus the plasma losses are ruled by the fast transported species: across the magnetic field lines this is the ions, and along the magnetic field lines it is the electrons. This effect is sometimes called the Simon short-circuit effect. [6]. It plays a decisive role in ion sources for multiply charged ions especially in the so-called ECRIS, a microwave discharge in a magnetic trap consisting of a superposition of a mirror trap as in Fig. 8 and a magnetic hexapole – see the discussion in Refs. [7], [8] and the literature cited therein. In these sources, it was found that biasing the metallic endplate of the discharge vessel greatly enhanced the production of high charge states. A further increase could be obtained by covering the inside of the discharge vessel with dielectric layers. Both measures intercept part of the compensating wall current and thus dramatically improve plasma containment. This in turn improves ionization into high charge states.

### 3.3   Fluid description of plasma

The discussion of single-particle motion neglects the strong coupling between positive and negative charges and thus cannot render all aspects of plasma physics. A model emphasizing more strongly the coupling of charges of opposite sign is the description of plasma as a conducting fluid. This description of plasma is called magneto fluid dynamics or plasma fluid dynamics.

In this model the fluid is considered as a continuum, which means that any partition of the fluid has the same properties – independently of its size. The granulation due to the atomic structure is neglected. This means that the discussion is only valid on macroscopic scales.

The kinematics of point masses describes the motion of a (rigid) body by its position vector $\vec{r}(t)$ and its time derivatives, velocity $\dot{\vec{r}}(t)$ and acceleration $\ddot{\vec{r}}(t)$. In fluid mechanics we have an extended medium, which we have to characterize by extended velocity and acceleration fields $\dot{\vec{r}}(t,x)$ and $\ddot{\vec{r}}(t,x)$. A position corresponds to a fluid particle, which should keep its identity while streaming. Thus we can define a trajectory for it. As a consequence, there exists an unambiguous transformation between the locations of fluid particles at a time $t_0$ and a later moment $t$. Let $\vec{r}(0) = \{x(0), y(0), z(0)\}$ be the coordinate of a particle at a time $t_0$, which we will in future denote by

$$\vec{r}(0) \equiv \vec{a} \equiv (\xi(0), \eta(0), \zeta(0)). \tag{46}$$

At a later time $t$ the position vector $\vec{r}$ of this particle is an unambiguous function of $\vec{a}$ and $t$:

$$\vec{r}(\vec{a},t) \equiv (x(\vec{a},t), y(\vec{a},t), z(\vec{a},t)). \tag{47}$$

This latter relation can be reversed and solved for $\vec{a}$ (at least in principle), yielding

$$\vec{a}(\vec{r},t) = (\xi(\vec{r},t), \eta(\vec{r},t), \zeta(\vec{r},t)). \tag{48}$$

Here $\vec{r}(\vec{a},t)$ is the position of the particle that at $t_0$ was at the position $\vec{a}$. The coordinate system of $\vec{r}(\vec{a},t)$ is called Euler coordinates. Their origin is fixed in space.

The coordinate $\vec{a}(\vec{r},t)$ identifies the fluid particle that at time $t$ is found at the position $\vec{r}$. The respective coordinate system is named Lagrange coordinates. These coordinates identify fluid particles and are thus coupled to the flow. Other names for Lagrange coordinates are convective or material coordinates.

A streaming continuum represents a velocity field. By this phrase we denote that the velocity is a function of position and time, that is

$$\vec{v} = \vec{v}(\vec{r},t) \quad \text{or} \quad \vec{v} = \vec{v}(\vec{a},t). \tag{49}$$

The two expressions can be unambiguously transformed into each other, but have different meanings: $\vec{v}(\vec{r},t)$ is the velocity of the fluid at the position $\vec{r}$ at the time *t*, while $\vec{v}(\vec{a},t)$ is the velocity of particle $\vec{a}$ at the time $t$. Besides velocity, a streaming fluid is also characterized by other quantities that are functions of position and space, like the mass density $\rho$, the pressure *p* and the temperature *T*. These quantities are called *intensive quantities*. By taking the volume integral over an intensive quantity, we end up with an *extensive quantity*. For example, the mass *m* of a certain area of the fluid is given by

$$m = \iiint_V \rho \, \mathrm{d}V. \tag{50}$$

Extensive quantities are additive, intensive quantities not: the mass of two regions is the sum of their masses. But for a homogeneous fluid, the mass density does not change, when adding material domains with equal densities together.

We now consider the space and time derivatives of intensive quantities. Consider such a quantity $\Phi(\vec{a}(\vec{r},t),t) = \Phi(\vec{r}(\vec{a},t),t)$. Here we understand that $\Phi$ is the *numerical value* of this function. In Euler coordinates we obtain for the differential $\mathrm{d}\Phi$

$$\mathrm{d}\Phi = \frac{\partial \Phi}{\partial t}\mathrm{d}t + \frac{\partial \Phi}{\partial x}\mathrm{d}x + \frac{\partial \Phi}{\partial y}\mathrm{d}y + \frac{\partial \Phi}{\partial z}\mathrm{d}z. \tag{51}$$

The convective time derivative is usually written $\mathrm{D}/\mathrm{D}t$. It is not a total derivative because $\vec{a}$ is kept constant. For the convective time derivative of $\Phi$ we obtain

$$\frac{\mathrm{D}\Phi}{\mathrm{D}t} = \left(\frac{\partial \Phi}{\partial t}\right)_{\vec{r}} + \frac{\partial x}{\partial t}\cdot\frac{\partial \Phi}{\partial x} + \frac{\partial y}{\partial t}\cdot\frac{\partial \Phi}{\partial y} + \frac{\partial z}{\partial t}\cdot\frac{\partial \Phi}{\partial z} = \frac{\partial \Phi}{\partial t} + (\vec{v}\cdot\vec{\nabla})\Phi. \tag{52}$$

This formula can be applied to the components of the vector $\vec{v}$, yielding

$$\frac{\mathrm{D}\vec{v}}{\mathrm{D}t} = \frac{\partial \vec{v}}{\partial t} + (\vec{v}\cdot\vec{\nabla})\vec{v}. \tag{53}$$

Note that the dot product on the right-hand side is a scalar, which multiplied with the vector $\vec{v}$ gives a vector. The term $\partial\vec{v}/\partial t$ is the so-called *local acceleration* of a non-stationary flow; the term $(\vec{v}\cdot\vec{\nabla})\vec{v}$ is named *convective acceleration*. To understand the difference, we consider the flow of an incompressible fluid in a tube with varying cross-section (e.g. a Venturi tube; see the sketch shown in Fig. 10). If the flow is stationary, the local acceleration is zero, $\partial\vec{v}/\partial t = 0$. However, a fluid particle passing from the left large cross-sectional area into the narrow one will be accelerated because the flow velocity is higher in the narrow section of the tube. This acceleration is described by the convective acceleration $(\vec{v}\cdot\vec{\nabla})\vec{v}$. If, however, the flow becomes non-stationary because the pressure at the left entrance changes, the flow velocity will change everywhere. Then $\partial\vec{v}/\partial t \neq 0$.

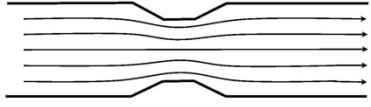

**Fig. 10:** Sketch of the stationary flow of an incompressible fluid in a tube with changing cross-sections

We now have the ingredients for formulating the equation of motion. We consider a fluid particle with mass $m$ given by Eq. (50). Here $\rho$ is the mass density of the fluid in the vicinity of the particle. Newton's equation for this particle reads

$$\frac{D}{Dt}\iiint_{\hat{V}} \rho\vec{v}\, dV = \sum \vec{F}. \tag{54}$$

Here $\sum \vec{F}$ is the summation over all forces acting on the particle. Our equation is in convective coordinates, the integration must be taken over $\hat{V}$, the volume of the fluid particle, which may change along the particle's trajectory. Thus the differentiation and the integration cannot simply be interchanged. In contrast to a volume $V$, fixed in space, we call $\hat{V}$ a *material domain*. However, for transforming Eq. (54) into Euler coordinates, we must replace the integration over $\hat{V}$ by integration over that fixed volume $V$, which just coincides with $\hat{V}$. The change of an extensive quantity of the particle inside this fixed volume is described by the *Reynolds transport theorem*, which states that the temporal change of an extensive quantity in a material domain is given by the temporal change of this quantity in the fixed volume just coinciding with the material domain and the total flow into and out of the fixed volume. Thus

$$\frac{D}{Dt}\iiint_{\hat{V}} \rho\vec{v}\, dV = \iiint_{V}\left(\frac{\partial \rho\vec{v}}{\partial t} + (\nabla \cdot \vec{v})\rho\vec{v}\right)dV. \tag{55}$$

Here and in what follows, we omit indicating the symbol of the nabla operator $\nabla$ by an arrow on top, as it is evident that it can be considered as a vector. Because $V$ is fixed, we can interchange time derivation and volume integration. To proceed further, we must specify the forces on the right-hand side of Eq. (54). We distinguish between *surface forces* and *volume forces*. Surface forces can be represented by a surface integral of a force density. The only example we have to consider is the pressure force $\vec{F}_p$ defined by

$$\vec{F}_p = -\iint_{A} p\, \vec{dA} = -\iiint_{V}(\nabla p)\, dV. \tag{56}$$

For the latter transformation, we used the Gauss theorem.

Volume or external forces $\vec{F}_V$ will be specified below. They can be represented by a volume integral over a force density $\vec{f}$. Thus

$$\vec{F}_V = \iiint_{V} \vec{f}\, dV. \tag{57}$$

By using these expressions we obtain for our momentum equation

$$\iiint_V \left( \rho \frac{\partial \vec{v}}{\partial t} + \rho (\vec{v} \cdot \nabla) \vec{v} + \nabla p - \vec{f} \right) dV = 0. \tag{58}$$

This equation must be valid for any fixed volume. This is only possible if the integrand equals zero:

$$\frac{\partial \vec{v}}{\partial t} + (\vec{v} \cdot \nabla) \vec{v} + \frac{1}{\rho} \nabla p - \frac{1}{\rho} \vec{f} = 0 \quad \text{(Euler equation)}. \tag{59}$$

Examples of external forces are the Lorentz force and the gravitational force. Further, we have an internal friction $\vec{f}_R$ due to momentum exchange between the different plasma components (and due to viscosity, but viscosity can be neglected in ion source plasma).

The Lorentz force acting on a single ion or electron is given by

$$q_k \vec{E} + q_k \vec{v}_k \times \vec{B} \equiv \frac{1}{n_k} \vec{f}_k. \tag{60}$$

Here the index *k* characterizes the particle species (electron or ions of different charge and mass). The total Lorentz force density is the sum over these different contributions:

$$\vec{f}_L = \sum \vec{f}_k = \left( \sum_k n_k q_k \right) \cdot \vec{E} + \left( \sum_k n_k q_k \vec{v}_k \right) \times \vec{B} = \rho_{el.} \vec{E} + \vec{j} \times \vec{B}. \tag{61}$$

Here we have introduced for abbreviation the electrical space-charge density $\rho_{el.} \equiv \sum_k n_k q_k$ and the electrical current density $\vec{j} \equiv \sum_k n_k q_k \vec{v}_k$. Analogously we obtain for the gravity force density ($\vec{g}$ is gravitational field strength)

$$\vec{f}_g = \rho \vec{g}. \tag{62}$$

Thus we have finally

$$\rho \frac{\partial \vec{v}}{\partial t} + \rho (\vec{v} \cdot \nabla) \vec{v} + \nabla p - \rho_{el.} \vec{E} - \vec{j} \times \vec{B} - \rho \vec{g} + \vec{f}_R = 0. \tag{63}$$

Without the Lorentz term, this equation is known as the *Navier–Stokes equation*. For some situations, it is sufficient to use this equation as a global plasma equation and consider ions and electrons to be strongly coupled. For instance, this may be the case for slow waves (model of plasma as a single fluid).

In the case of high-frequency waves, the ion inertia may be so large that ions cannot follow the fast oscillations, while electrons can. In this case it is better to formulate a separate fluid equation for every plasma component and consider the coupling by mutual friction and electric fields created by space charges. Neglecting gravitation, we have for a component *k*

$$\rho_k \frac{\partial \vec{v}_k}{\partial t} + \rho_k (\vec{v}_k \cdot \nabla) \vec{v}_k + \nabla p_k - \rho_{el.,k} \vec{E} - \vec{j}_k \times \vec{B} + \sum_l \vec{f}_{k,l} = 0. \tag{64}$$

The last term contains the forces due to internal friction between the different plasma species, which is between electrons or ions and electrons, or between ions and neutrals. A *highly ionized plasma* is defined as plasma where electron–ion friction dominates. A *weakly ionized plasma* is dominated by friction between charged particles and neutrals.

For the force density $\vec{f}_{k,l}$ acting on particles of kind $k$ due to collisions with particles of kind $l$, we have the general formula

$$\vec{f}_{kl} = n_k m_{kl} \nu_{kl} (\vec{v}_l - \vec{v}_k) , \qquad (65)$$

where

$$m_{kl} = \frac{m_k m_l}{m_k + m_l} \qquad (66)$$

is the reduced mass of the colliding particles and $\nu_{kl}$ is the respective collision frequency for momentum transfer.

In addition, we need the continuity equation describing the conservation of mass

$$\frac{\partial \rho}{\partial t} + (\vec{v} \cdot \nabla)\rho + \rho \nabla \cdot \vec{v} = 0 , \qquad (67)$$

which can also be formulated for each of the different plasma components, and what is called Ohm's law, but is in reality the definition of the electrical conductivity $\sigma$. (Ohm's law states that $\sigma$ depends neither on the electric field $\vec{E}$ nor on the current density $\vec{j}$; only in this we find the case $\vec{j} \propto \vec{E}$.) The current density in a flowing fluid is proportional to the electric field $\vec{\tilde{E}}$ in the frame of the moving fluid. Transformation to a system fixed in space yields $\vec{\tilde{E}} = \vec{E} + \vec{v} \times \vec{B}$. Thus the current density is proportional to $\vec{E} + \vec{v} \times \vec{B}$ in a coordinate system at rest. The proportionality constant is the conductivity $\sigma$:

$$\vec{j} = \sigma(\vec{E} + \vec{v} \times \vec{B}) . \qquad (68)$$

In the presence of an external magnetic field, plasma is anisotropic. The density of electric currents flowing under the action of an electric field $\vec{E}$ is not necessarily parallel to $\vec{E}$. Thus $\sigma$ must be considered as a tensor in the general case. We express this by doubly underlining the symbol. The more general definition of the conductivity thus reads

$$\vec{j} = \underline{\underline{\sigma}}(\vec{E} + \vec{v} \times \vec{B}) . \qquad (69)$$

We discuss this in detail in the following section. Now

$$\vec{j} = e n_i \vec{v}_i - e n_e \vec{v}_e . \qquad (70)$$

The ion and electron densities must be obtained from a two-fluid model.

## 3.4 AC conductivity of magnetized plasma

We now consider the case of particle motion under the action of an oscillating electric field $\vec{E} = \vec{E_0} \exp(-i\omega t)$ in magnetized plasma. The vectors $\vec{E_0}$ and $\vec{B}$ define a plane. We introduce Cartesian coordinates (main directions labelled $x, y, z$) in such a way that this plane is the $xz$-plane and $\vec{B}$ is parallel to the $z$-direction (see Fig. 11).

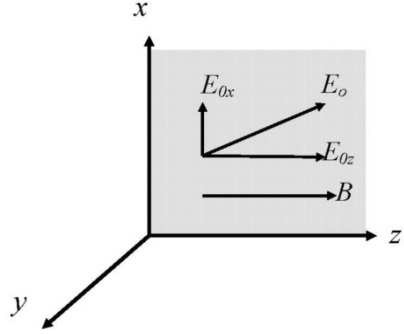

**Fig. 11:** Coordinates for calculating the AC conductivity of magnetized plasma

To solve the equation of motion of a charged particle, we start with an ansatz for the particle velocity $v$

$$\vec{v} = \left[ a_1(t)\vec{E_{01}} + a_2(t)\vec{E_{01}} \times \vec{B} + a_3(t)\vec{E_{03}} \right] \exp(-i\omega t). \tag{71}$$

(Note that $\vec{E_0}$ has by definition no component in the $y$-direction. We therefore constructed a respective coordinate vector by the cross product between $\vec{E_{01}}$ and $\vec{B}$). Introducing Eq. (71) into Eq. (33) gives, after some algebra, a set of three equations for the velocity components, respectively for the $a_i$. Of these equations, those for $a_{1,2}$ are coupled, while the equation for $a_3$ does not depend on the magnetic induction $\vec{B}$, and is similar to the equation for the case without a magnetic field, namely

$$\frac{\partial a_3}{\partial t} - i\omega a_3 - \frac{q}{m} = 0 \quad \Rightarrow \quad a_3 = \frac{iq}{\omega m} \quad \Rightarrow \quad v_z = \frac{iqE_z}{\omega m}. \tag{72}$$

For $a_{1,2}$ we obtain

$$\frac{\partial a_1}{\partial t} - i\omega a_1 - \frac{q}{m} + \frac{qB^2 a_2}{m} = 0,$$

$$\frac{\partial a_2}{\partial t} - i\omega a_2 - \frac{qa_1}{m} = 0. \tag{73}$$

For separating and solving (73), we differentiate both equations and in the set of differentiated equations we replace the first derivatives of the $a_i$ by the expressions of (73). By using (37) and with some algebra we obtain for $a_2$ the inhomogeneous differential equation

$$\frac{\partial^2 a_2}{\partial t^2} - 2i\omega \frac{\partial a_2}{\partial t} + (\omega_B^2 - \omega^2) a_2 = \frac{q^2}{m^2}. \tag{74}$$

The solution of the homogenous part of (74) is the gyration of the particle and will not be considered here. A special solution of the inhomogeneous equation yields for $\omega \neq \omega_B$

$$a_2 = \frac{q^2}{m^2} \frac{1}{\omega_B^2 - \omega^2} \quad \text{and} \quad a_1 = \frac{q}{m} \frac{i\omega}{\omega_B^2 - \omega^2}. \tag{75}$$

These solutions correspond to forced oscillations with a resonance at $\omega = \omega_B$. In the case of this resonance, one obtains a different solution describing a gyration with the amplitude rising linearly in time. This is the well-known cyclotron resonance, where a charged particle gains energy continuously from the electric field.

A particle moving under the action of a constant force $\vec{F}$ in a viscous medium attains after some time a finite, constant velocity $\vec{v}$, which is proportional to $\vec{F}$. The proportionality constant is called the mobility $b$. Thus we have $\vec{v} = b\vec{F}$. The non-resonant oscillatory motion of a charged particle under the action of an oscillating force has constant amplitude proportional to the amplitude of the oscillating field even if there is no viscous medium. This is due to the inertia of the oscillating particles creating a phase shift of 90° between the force and the particle velocity. This limits the gain of energy in the oscillating electric field. Thus also in this case the proportionality constant between the amplitude of the force and the amplitude of the velocity is called the mobility, because of some analogy with the DC case. In complex notation, assuming an electric field of the form $\vec{E} = \vec{E}_0 \exp(-i\omega t)$, we obtain from the equation of motion

$$\vec{v} = \frac{iq}{\omega m} \vec{E}. \tag{76}$$

Thus the mobility is given by $b = iq/\omega m$, the imaginary unit i standing for a phase shift of $\pi/2$ between the velocity and the driving force $q\vec{E}$.

To generalize the case of magnetized plasma considered above, we assume the existence of an oscillating electric field $\vec{E} = \vec{E}_0 \exp(-i\omega t) = \vec{E}_1 + \vec{E}_2 + \vec{E}_3$. Here the $\vec{E}_i$ are the components of $\vec{E}$ in the coordinate directions. From the solution found above, we can construct the drift velocity $\vec{v}_{\text{drift}}$ of a charged particle under the action of this field,

$$\vec{v}_{\text{drift}} = \frac{i\omega}{\omega_B^2 - \omega^2} \frac{q}{m} (\vec{E}_1 + \vec{E}_2) + \frac{q^2/m^2}{\omega_B^2 - \omega^2} (\vec{E}_2 \times \vec{B} + \vec{E}_1 \times \vec{B}) + \frac{iq}{\omega m} \vec{E}_3, \tag{77}$$

corresponding to the component equations

$$\begin{aligned} v_{\text{drift},x} &= \frac{i\omega}{\omega_B^2 - \omega^2} \frac{q}{m} E_1 + \frac{q^2 B/m^2}{\omega_B^2 - \omega^2} E_2 + 0 \\ &\equiv b_{11} q E_1 + b_{12} q E_2 + b_{13} q E_3, \end{aligned}$$

$$\begin{aligned} v_{\text{drift},y} &= -\frac{q^2 B/m^2}{\omega_B^2 - \omega^2} E_1 + \frac{i\omega}{\omega_B^2 - \omega^2} \frac{q}{m} E_2 + 0 \\ &\equiv b_{21} q E_1 + b_{22} q E_2 + b_{23} q E_3, \end{aligned} \tag{78}$$

$$v_{\text{drift},z} = 0 + 0 + \frac{iq}{\omega m} E_3$$
$$\equiv b_{31} q E_1 + b_{32} q E_2 + b_{33} q E_3.$$

In the respective second lines, we have formulated the component equations with the components $b_{ij}$ of the mobility tensor $\underline{\underline{b}}$, which has to replace the scalar mobility $b$ in the case of the anisotropic magnetized plasma. From Eqs. (78) we obtain for $\underline{\underline{b}}$

$$\underline{\underline{b}} = \frac{i}{\omega m} \begin{Vmatrix} \dfrac{\omega^2}{\omega^2 - \omega_B^2} & \dfrac{\mp i\omega\omega_B}{\omega^2 - \omega_B^2} & 0 \\ \dfrac{\pm \omega\omega_B}{\omega^2 - \omega_B^2} & \dfrac{\omega^2}{\omega^2 - \omega_B^2} & 0 \\ 0 & 0 & 1 \end{Vmatrix} \equiv \frac{i}{\omega m} \underline{\underline{K}}. \tag{79}$$

Having found the mobility, we can define the conductivity of a medium. If we have drifting plasma, the density $\vec{j}$ of an electric current is given as the sum over the densities of the drift currents of the charged particles identified by the index $k$. In an isotropic medium (plasma without external magnetic field) we have

$$\vec{j} = \sum_k n_k q k \vec{v}_{\text{drift},k} = \sum_k n_k q_k b_k q_k \vec{E} \equiv \sigma \vec{E}. \tag{80}$$

The direction of $\vec{j}$ depends only on the direction of $\vec{E}$, not on the sign of the charge $q$. From (80) we obtain

$$\sigma = \sum_k n_k b_k q_k^2. \tag{81}$$

The sign of the conductivity does not depend on the sign of $q$. In the case of an anisotropic medium, we obtain analogously

$$\underline{\underline{\sigma}} = \sum_k n_k \underline{\underline{b}}_k q_k^2 \quad \Rightarrow \quad \sigma_{ij} = \sum_k n_k (b_{ij})_k q_k^2. \tag{82}$$

Thus by using (79) we have

$$\underline{\underline{\sigma}}(\omega) = \frac{i}{\omega} \sum_k \frac{n_k q_k^2}{m_k} \underline{\underline{K}}_k. \tag{83}$$

Here $\underline{\underline{K}}_k$ is the tensor defined in Eq. (79) for the plasma component labelled $k$.

What we have considered here is collision-free plasma. In this case the plasma is a loss-free reactance and $\underline{\underline{\sigma}}$ is imaginary, that is, without ohmic contributions. In the case of collisions, we have to supplement Eq. (33) by the average momentum loss due to collisions

$$\delta \vec{p}_k / \delta t = \sum_{j \neq k} m_{rjk} (\vec{v}_j - \vec{v}_k) \nu_m (\vec{v}_j - \vec{v}_k). \tag{84}$$

Here the $m_{rjk}$ are the reduced masses of the colliding particles, the $\vec{v}_{j,k}$ are the particle velocities and $v_m(|\vec{v}_j - \vec{v}_k|)$ is the average momentum transfer collision frequency. By adding this term, the elements of the conductivity tensor will become complex, the real parts describing the ohmic losses in the plasma.

## 4  Plasma waves

### 4.1  Some general wave concepts

Let us recall the concept of electromagnetic waves in vacuum as described by Maxwell's equations, which read

$$\nabla \times \vec{H} = \varepsilon_0 \frac{\partial \vec{E}}{\partial t} \quad (\vec{j} = 0, \text{ no electric current}),$$
$$\nabla \times \vec{E} = -\mu \frac{\partial \vec{H}}{\partial t} \quad (\nabla \cdot \vec{E} = 0, \text{ no space charge}, \nabla \cdot \vec{B} = 0, \text{ no magnetic monopoles}).$$
(85)

Multiplying the first equation by (the negative of) the vacuum permittivity ($-\mu_0$), and taking the time derivative yields

$$-\mu_0 \nabla \times \frac{\partial \vec{H}}{\partial t} = -\frac{1}{c^2} \frac{\partial^2 \vec{E}}{\partial t^2},$$
(86)

where we have used the relation $\varepsilon_0 \mu_0 = 1/c^2$, and $c$ is the phase velocity of light in vacuum. Taking the curl ($\nabla \times$) of the second equation and eliminating $\vec{H}$ by using Eq. (85??), we finally get (in similar ways)

$$\left(\nabla^2 - \frac{1}{c^2} \frac{\partial^2}{\partial t^2}\right) \vec{E} = 0 \quad \text{and} \quad \left(\nabla^2 - \frac{1}{c^2} \frac{\partial^2}{\partial t^2}\right) \vec{H} = 0.$$
(87)

These are the equations for electromagnetic waves in vacuum. We solve them by assuming plane waves

$$\vec{E} = \vec{E_0} \exp[i(\vec{k} \cdot \vec{r} - \omega t)] + \vec{E_0}^* \exp[-i(\vec{k} \cdot \vec{r} - \omega t)].$$
(88)

Here $\vec{E_0}$ and $\vec{E_0}^*$ are the complex amplitudes, the wave vector $\vec{k}$ gives the direction of wave propagation, and $\omega$ is the angular frequency. The asterisk indicates the complex conjugate. When introducing this ansatz into the wave equation, we obtain equations in which $\nabla$ is replaced by $\pm i\vec{k}$ and the time derivative by $\mp i\omega$. In principle, one needs to take only one of the two terms on the right-hand side of (88), because only the real part is physically important. It is the same for both terms. We will take for our discussion the first term (i.e. the upper sign), but you may find a different convention in some books.

In the case of vacuum we obtain

$$-k^2 + \omega^2 c^2 = 0 \quad \text{or} \quad \omega = \pm kc.$$
(89)

In general, the relations $\omega(k)$ or $k(\omega)$ are termed *dispersion relations*. For vacuum, the dispersion relation is linear.

We now consider the argument of the exponential functions, that is, the *phase*. By the condition

$$\frac{d}{dt}(\vec{k} \cdot \vec{r} - \omega t) = \vec{k} \cdot \dot{\vec{r}} - \omega = 0 \tag{90}$$

we define a velocity for the propagation of the phase. It follows from this equation that the phase travels parallel to $\vec{k}$ and that the absolute value of the phase velocity is given by

$$v_{ph} = \frac{\omega}{k}. \tag{91}$$

By

$$\mu = c/v_{ph} = c \cdot k/\omega \tag{92}$$

we define the refractive index of a wave. In vacuum we have $c = v_{ph}$, thus $\mu = 1$.

Besides the phase velocity we define the group velocity $v_g$ by

$$v_g = \frac{d\omega}{dk} = \frac{d(v_{ph}k)}{dk} = v_{ph} + k\frac{dv_{ph}}{dk} = v_{ph} - \lambda\frac{dv_{ph}}{d\lambda}. \tag{93}$$

## 4.2 Waves in plasma without external magnetic field

### 4.2.1 Electromagnetic waves

As a first step we consider waves in plasma without an external magnetic field, that is, for $\vec{B} = 0$. The convective acceleration is *per se* nonlinear. By neglecting this term, we restrict our consideration to waves with sufficiently small amplitudes as a first step. In contrast to vacuum, we may have space charge and electric current in plasma. Thus $\nabla \cdot \vec{E} \neq 0$ and $\vec{j} = \rho_{el}\vec{v} \neq 0$.

However, we neglect pressure effects and assume the plasma to contain only singly charged ions of one kind. The equation of motion thus becomes

$$\frac{\partial \vec{j}}{\partial t} = e^2 n \left(\frac{1}{m_e} + \frac{1}{m_i}\right)\vec{E} = \frac{e^2 n}{m_{ei}}\vec{E} = \varepsilon_0 \omega_p^2 \vec{E}, \tag{94}$$

where $n = n_e = n_i$ is the plasma density, and $m_{ei} \approx m_e$ is the reduced mass of electrons and ions. Assuming plane waves yields

$$\vec{j} = -\frac{\varepsilon_0 \omega_p^2}{i\omega}\vec{E}. \tag{95}$$

Thus we obtain $\sigma = i\varepsilon_0\omega_p^2/\omega$. In this approximation, plasma is not an ohmic conductor but a reactance. There is a phase shift by 90° between the electric field and the current density. The imaginary conductivity is due to the inertia of the electrons. Plasma behaves like an inductance. The Maxwell equations for this case become

$$\nabla \times \vec{H} = -\frac{\varepsilon_0 \omega_p^2}{i\omega} \vec{E} + \varepsilon_0 \frac{\partial \vec{E}}{\partial t},$$

$$\nabla \times \nabla \times \vec{E} = -\frac{\omega_p^2}{ic^2\omega} \frac{\partial \vec{E}}{\partial t} - \frac{\omega^2}{c^2} \frac{\partial^2 \vec{E}}{\partial t^2}.. \tag{96}$$

Transformations of these equations must allow for transverse as well as longitudinal waves as possible solutions. Multiplying the first of these equations by $(-\mu_0)$ and with $\vec{E} = \vec{E_0}\exp[i(\vec{k}\cdot\vec{r} - \omega t)]$ we thus obtain finally

$$(\vec{k}\vec{k} - k^2)\vec{E} = \frac{\omega^2}{c^2}\left(1 - \frac{\omega_p^2}{\omega^2}\right)\vec{E} = \varepsilon \frac{\omega^2}{c^2}\vec{E}, \tag{97}$$

where

$$\varepsilon = 1 - \omega_p^2/\omega^2 \tag{98}$$

is the permittivity of the plasma. Equation (98) is known as the *Eccles relation*.

For transverse waves, $\vec{k} \perp \vec{E}$, thus $\vec{k}\cdot\vec{E} = 0$ and we obtain as dispersion relation

$$k^2 = \frac{\omega^2}{c^2}\left(1 - \frac{\omega_p^2}{\omega^2}\right) \quad \text{or} \quad \frac{\omega^2}{\omega_p^2} = 1 + \frac{k^2 c^2}{\omega_p^2}. \tag{99}$$

The latter form we present in Fig. 12 (so-called *Brillouin diagram*). Here the inclinations of the dashed (green) lines give the phase velocity and the group velocity of the plasma wave at the point where these lines cross the (blue) plasma wave curve.

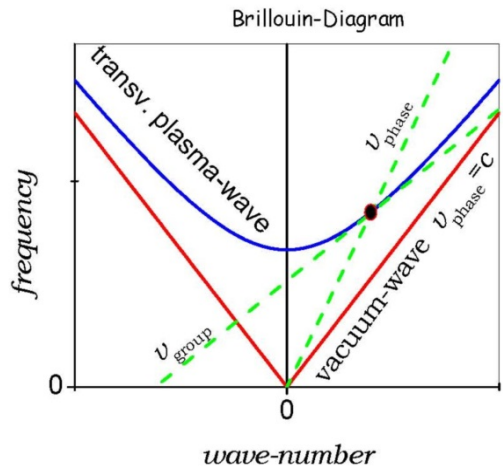

**Fig. 12:** Brillouin diagram (i.e. frequency versus wavenumber) for transverse electromagnetic waves in vacuum and in magnetic field-free plasma.

For longitudinal waves, $\vec{k} \parallel \vec{E}$ and thus $\vec{k}(\vec{k}\cdot\vec{E}) = k^2 \vec{E}$. The left-hand side of Eq. (97) equals zero. Thus

$$1 - \frac{\omega_p^2}{\omega^2} = 0 \quad \Rightarrow \quad \omega = \omega_p. \tag{100}$$

There is no wave, only oscillations, with the plasma frequency as discussed in the introduction.

For high frequencies, the dispersion relation of transverse waves approaches asymptotically that of free space. The plasma behaves as a dielectric with a refractive index $\mu$ given by the so-called Maxwell relation

$$\mu = \sqrt{\varepsilon} = \sqrt{1 - \omega_p^2/\omega^2} \leq 1. \tag{101}$$

The approximations used for deriving Eq. (101) become valid at sufficiently high frequencies of the waves. Even in the case of plasma with external magnetic field, our model is correct for $\omega \gg \omega_B$. We expect deviations at lower frequencies.

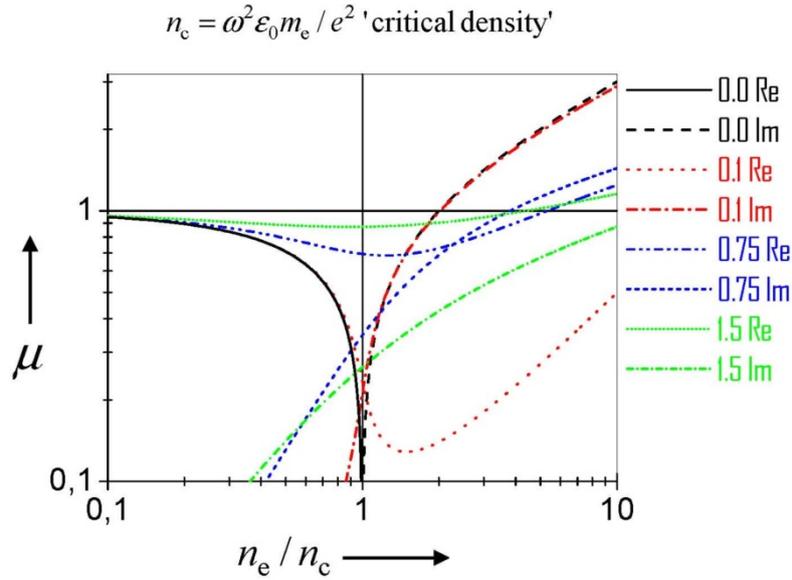

**Fig. 13:** Real (Re) and imaginary (Im) parts of the refractive index of transverse plasma waves versus plasma density over critical density for different ratios of momentum transfer collision frequency over wave frequency.

By using the refractive index with $\nu_m = 0$ we can distinguish different regions for transverse wave propagation in plasma (see Fig. 13):

– For $\omega > \omega_p$, the refractive index is real. Waves can propagate without damping. Plasma behaves as a lossless dielectric with a refractive index $\mu < 1$. Thus the phase velocity of waves exceeds the phase velocity of electromagnetic waves in vacuum, $v_{ph} > c$.

– For $\omega = \omega_p$, we have $\mu = 0$ and $k = 0$. There is no wave, only an oscillation.

- For $\omega < \omega_p$, both $\mu$ and $k$ are imaginary. There is no wave propagation. Waves when penetrating a plasma will decay exponentially over the so-called skin depth $L_{skin}$ given by

$$L_{skin} = \frac{1}{\text{Im}\, k} = \frac{1}{(\omega/c)\sqrt{\omega_p^2/\omega^2 - 1}} \quad \Rightarrow \quad \frac{c}{\omega_p} = \frac{\lambda_p}{2\pi} \quad \text{for} \quad \omega \to 0. \tag{102}$$

For small frequencies the skin depth does not depend on frequency and is equal to the wavelength of an electromagnetic wave in vacuum having a frequency equal to the plasma frequency.

Collisions between charged particles and between charged particles and neutrals cause wave damping. Since electronic collisions are most frequent, they are most important. When considering these collisions, instead of Eq. (94) we obtain

$$\frac{\partial \vec{j}}{\partial t} + \vec{j} \cdot \nu_m = \varepsilon_0 \omega_p^2 \vec{E}. \tag{103}$$

Introducing plane waves yields

$$\vec{j} = -\varepsilon_0 \omega_p^2 \frac{\vec{E}}{i\omega - \nu_m} \equiv \sigma \vec{E}. \tag{104}$$

By a similar procedure as described above, for the dispersion relation with respect to the refractive index we obtain

$$\mu = \frac{kc}{\omega} = \sqrt{\left(1 - \frac{\omega_p^2}{\omega^2 + \nu_m^2}\left(1 - i\frac{\nu_m}{\omega}\right)\right)}. \tag{105}$$

This function is plotted in Fig. 13 for different values of the collision frequency $\nu_m$. As a result of dissipation by collisions, the conductivity will become complex. The same holds true for the refractive index and for $k$. There are no longer frequency regions where $\mu$ is purely imaginary or real. Waves propagate even for $\omega < \omega_p$, though they are heavily damped. But damping is also observed for $\omega > \omega_p$.

### 4.2.2 Longitudinal waves

Longitudinal waves in neutral gases are (compressive) sound waves. In the equation of motion, the respective restoring force is described by the $\nabla p$ term. Acoustic waves are dispersion-free, that is, $\omega = kc_s$. The phase velocity of sound in neutral gas, $c_s$, is given by

$$c_s = \sqrt{\kappa p/\rho}. \tag{106}$$

Here $\kappa$ is the ratio of the specific heat at constant pressure to that at constant volume. To obtain these relations, one considers small wave amplitudes and neglects all terms quadratic in wave amplitude or quantities proportional to wave amplitudes (linearization). The gas is compressed adiabatically by the wave, thus $\nabla p/p = \kappa(\nabla n/n)$.

In analogy to Eq. (106), for abbreviation, in plasmas we can define two further sound speeds:

$$c_{si} = \sqrt{\kappa_i p_i / \rho_i} \quad \text{ion sound speed} \tag{107}$$

and

$$c_{se} = \sqrt{\kappa_e p_i / \rho_e} \quad \text{electron sound speed} . \tag{108}$$

For simplicity we neglect collisional damping, and obtain the equations of motion for electrons and ions as

$$\rho_e \frac{\partial \vec{v}_e}{\partial t} = -n_e e \vec{E} - \nabla p_e,$$
$$\rho_i \frac{\partial \vec{v}_i}{\partial t} = n_i e \vec{E} - \nabla p_i. \tag{109}$$

Here $e$ is the *positive* elementary charge. By a linearization similar to that applied for ordinary sound waves, considering longitudinal waves with $\vec{k} \parallel \vec{E}$ (this is equivalent to reducing the Maxwell equations to the Poisson equation) and assuming plane waves, one ends up with two combined equations of motion for electrons and ions. These equations have a non-trivial solution only if their determinant is zero. This condition yields the dispersion relation, a bi-quadratic equation having two independent solutions corresponding to two different kinds of wave. At high frequencies we obtain approximately

$$\omega^2 = \omega_p^2 + k^2(c_{se}^2 + c_{si}^2) \approx \omega_{pe}^2 + k^2 \kappa_e \frac{k_B T_e}{m_e} \quad \text{(Bohm–Gross dispersion relation)}. \tag{110}$$

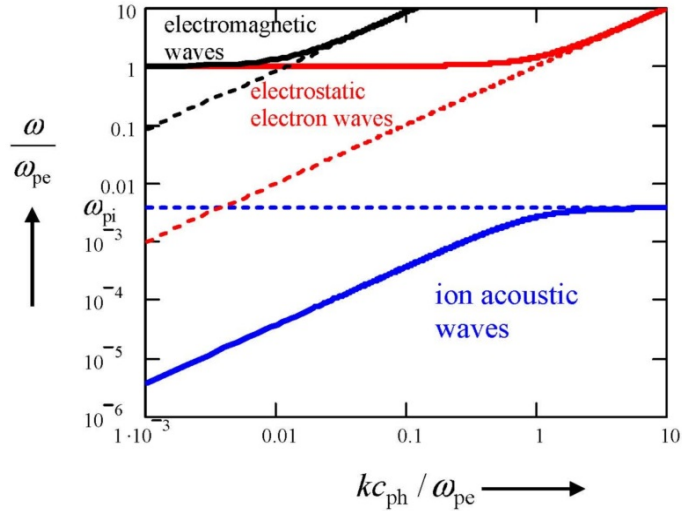

**Fig. 14:** Double-log Brillouin diagram for longitudinal plasma waves plotted from the exact formulas; and the same for electromagnetic (transverse) waves for comparison.

The respective waves are electrostatic electron waves. The dispersion relation is formally equivalent to that of electromagnetic waves. However, the phase velocity is much smaller in the case of acoustic waves: $c_{se} \ll c$. In Fig. 14 we show these waves for the case of argon plasma with cold ions, $n_e = 10^{18}\,\text{m}^{-3}$ and $k_B T_e = 10\,\text{eV}$. The coefficient $\kappa_e$ depends on the model used. Many authors set $\kappa_e = 3$ (see e.g. [9]).

The second branch represents ion acoustic waves. For small wavenumbers one obtains for the dispersion relation

$$\omega^2 \approx k^2 \left( c_{si}^2 + \frac{m_e}{m_i} c_{se}^2 \right) \approx k^2 \kappa_e \frac{k_B T_e}{m_i}. \tag{111}$$

The latter transformation is valid if (as mostly the case in low-pressure gas discharges) $T_i \ll T_e$. For larger values of $k$ (i.e. at short wavelengths), the frequency is almost constant and corresponds to the ion plasma frequency $\omega_{pi}$ as sketched in Fig. 14 for an argon plasma with the values of density and electron temperature given above. We present the exact dispersion relation, not the approximations given above. To span the large differences in frequency of these types of waves, we have used a double-logarithmic plot (compare the Brillouin diagram, Fig. 12).

In reality electrostatic waves are strongly damped, especially at short wavelengths. The damping mechanism is not described by the fluid dynamic model used here. It relies on the concept of fluid particles. In a plasma any set of microscopic particles (electrons and ions) forming a (macroscopic) fluid particle will disintegrate by diffusion within a short time. Thus ions and electrons come out of phase with the wave motion and their wave energy is transformed into thermal energy. Further, the interaction of ions and electrons with the electric field of a wave depends strongly on the particle velocity and is most efficient for particles moving in the same direction as the wave with velocities close to the phase velocity. The waves are damped by transferring energy to those particles. These nonlinear processes (*Landau damping*, *particle trapping*) are adequately described only by kinetic theory.

## 4.3 Waves in plasma with external magnetic field

### 4.3.1 Cold plasma

When plasma is magnetized, not only does it become anisotropic, but also the force exerted by the magnetic field on charged particles will increase the number of possible wave modes. To describe all this would need more than the space available here. Therefore, we will discuss only a simple example as a model and then give some general remarks. To start, we neglect the $\nabla p$ force. This corresponds to considering only cold, homogeneous plasma. Our set of equations is the appropriate Maxwell equation,

$$\nabla \times \nabla \times \vec{E} = \nabla^2 \vec{E} - \nabla(\nabla \cdot \vec{E}) = \frac{1}{c^2} \frac{\partial^2 \vec{E}}{\partial t^2} + \frac{1}{\varepsilon_0 c^2} \frac{\partial \vec{j}}{\partial t}, \tag{112}$$

the momentum equation from a single-fluid model,

$$\rho \frac{D\vec{v}}{Dt} = \rho_{el} \vec{E} + \vec{j} \times \vec{B}, \tag{113}$$

and (a simplified form of) the so-called Ohm's law obtained from a two-fluid model,

$$\frac{\partial \vec{j}}{\partial t} = \frac{e^2 n}{m_e} (\vec{E} + \vec{v} \times \vec{B}) - \frac{\omega_{Be}}{B} (\vec{j} \times \vec{B}). \tag{114}$$

The decisive difference, that is, the anisotropy of the plasma, is represented by the $\vec{j} \times \vec{B}$ terms. As a consequence, $\vec{j}$ and $\vec{E}$ may not be parallel. For linearization we use the scheme

$$\vec{B} = \vec{B}_0 + \vec{B}', \quad \vec{E} = 0 + \vec{E}', \quad \vec{j} = 0 + \vec{j}', \quad \vec{v} = 0 + \vec{v}', \quad \rho = \rho_0 + \rho', \quad \rho_{\mathrm{el}} = 0 + \rho'_{\mathrm{el}}. \tag{115}$$

The primed quantities are considered as small perturbations of the equilibrium values and small to first order. When introducing (115) into Eqs. (112)–(114) and neglecting terms with products of small quantities, because they are small to second order, we obtain a new system of linear equations. By setting the determinant of this equation system to zero, we obtain a functional $F(\omega, k)$ as dispersion relation.

We simplify this procedure for two special cases, assuming transverse waves with $\vec{E} \perp \vec{k}$, propagating along the magnetic induction of the external field, that is, $\vec{k} \parallel \vec{B}_0$. By eliminating the electric field, we obtain an equation for $\vec{j}$,

$$\left\{ \omega^2 - \frac{\omega^2 \omega_{\mathrm{p}}^2}{\omega^2 - k^2 c^2} - \omega_{\mathrm{Bi}} \omega_{\mathrm{Be}} \right\} \vec{j} = -\mathrm{i} \omega \omega_{\mathrm{Be}} \vec{j} \times \vec{B}_0. \tag{116}$$

Here $\omega_{\mathrm{Bi,Be}}$ are the gyro-frequencies of ions and electrons respectively (see Eq. (37)). For $\omega \ll \omega_{\mathrm{Bi}}$ we can neglect the term quadratic in $\omega$ and the term on the right-hand side of Eq. (116) and obtain for the refractive index of these waves

$$\mu^2 = \left(\frac{kc}{\omega}\right)^2 = 1 + \frac{nm_{\mathrm{i}}}{\varepsilon_0 B_0^2} \approx \frac{\rho \mu_0 c^2}{B^2} \equiv \frac{c^2}{v_{\mathrm{A}}^2}. \tag{117}$$

The phase velocity of these low-frequency waves is the so-called Alfvén velocity $v_{\mathrm{A}}$. They were first described by H. Alfvén and are called Alfvén waves. These waves resemble waves on a string under tension. Here the oscillating strings are the plasma-filled magnetic flux tubes.

If we multiply Eq. (116) with its complex conjugate, we obtain an equation that can be reduced by erasing $|\underline{j}|^2$ on both sides. Taking the square root and after some algebra we obtain the refractive index

$$\mu^2 = 1 - \frac{\omega_{\mathrm{p}}^2}{\omega^2 - \omega_{\mathrm{Bi}} \omega_{\mathrm{Be}} \mp \omega \omega_{\mathrm{Be}}}. \tag{118}$$

We see that $\mu^2 \to \infty$ and changes sign for

$$\omega^2 - \omega_{\mathrm{Bi}} \omega_{\mathrm{Be}} \mp \omega \omega_{\mathrm{Be}} = 0. \tag{119}$$

The two signs in the above correspond to two different types of waves, which show at two specific frequencies a resonance with the refractive index growing over all limits. For $\omega \gg \omega_{\mathrm{Bi}}$ we can neglect the term $\omega_{\mathrm{Bi}} \omega_{\mathrm{Be}}$ and obtain as resonance frequency $\omega = \omega_{\mathrm{Be}}$. For $\omega \ll \omega_{\mathrm{Be}}$ we can neglect the $\omega^2$ term and obtain as resonance frequency $\omega = \omega_{\mathrm{Bi}}$. (The values of the resonance frequencies are exact despite the many approximations made.) These are the electron cyclotron and ion cyclotron waves. Both are circularly polarized: electron cyclotron waves right-hand, and ion cyclotron waves left-hand. The electric field vector changes its direction like the velocities of the gyrating electrons and ions, respectively. At resonance, electrons and ions respectively gyrate synchronously with the wave and can extract energy continuously.

Where the refractive index is imaginary, waves cannot propagate but are damped. The plasma behaves like a waveguide with propagation and cut-off regions (respectively stop bands). The frequencies limiting these regions are either the resonances ($|\mu^2|\to\infty$) or the cut-off frequencies where $\mu=0$. The correlation between resonances, cut-offs and stop bands is shown schematically in Fig. 15. Stop bands are shaded. In non-magnetized plasma we find only a cut-off frequency, the plasma frequency, and no resonance (see Fig. 13). There we also demonstrated the effect of damping. As a consequence, $\mu^2$ will be a complex number, its value remaining finite and non-zero, resonances and cut-offs becoming diffuse. But the cut-off regions remain regions of very strong damping, while damping in the propagation regions is much weaker. The simple damping-free description gives still the main features.

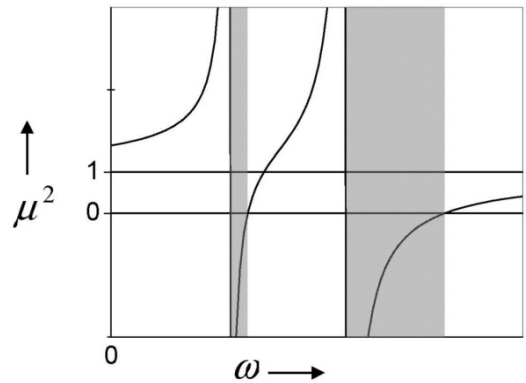

**Fig. 15:** The square of the refractive index $\mu^2$ versus (angular) wave frequency $\omega$. The shaded regions define stop bands with imaginary refractive index (schematic).

In magnetized plasma, waves may propagate in any direction; however, wave properties like refractive index (phase velocity) and resonance frequencies depend on the direction of propagation. An exception is the cut-off frequencies. They are the same for all directions. Usually one only considers wave propagation in the so-called principal directions: along and across the external magnetic field. Waves are classified according to their properties in the main directions. But these classifications may lose their sense when considering oblique wave propagation.

For cold infinite plasma, the dispersion relation again has two solutions, describing fast and slow wave types. These waves have different names depending on the direction of propagation, frequency range and also on the author. For waves where the electric vector $\vec{E}$ is parallel to the field lines of the external magnetic field (respectively the induction $\vec{B}_0$), the dispersion relation is the same as in the case of non-magnetized plasma. These waves are called ordinary waves. Thus this name is used in plasma physics in a manner different from in crystal optics (where 'ordinary' means waves propagating in any direction with the same phase speed). In plasma, ordinary waves exist only in the so-called main directions of wave propagation: $\vec{k}\parallel\vec{B}_0$ and $\vec{k}\perp\vec{B}_0$. For $\vec{k}\parallel\vec{B}_0$, the condition is fulfilled only for longitudinal waves; for $\vec{k}\perp\vec{B}_0$, only for transverse waves. Ordinary plasma waves propagating in other directions do not exist, that is, the dispersion relations for waves propagating in any other direction differ from that of the magnetic field-free case. In warm plasma, where pressure effects are important, at least two more electrostatic wave types can exist.

In bounded plasma there is a limitation by wavelength. Long-wavelength waves cannot propagate. On surfaces, special types of surface waves are possible. How does one find orientation? For the infinite plasma we will give a general treatment ending in the very helpful Clemmow–Mullaly–Allis (CMA) diagram.

### 4.3.2 General dispersion relation for cold magnetized plasma

For cold magnetized plasma, the Maxwell equation for plane waves reads

$$i\vec{k} \times \vec{H} = \vec{j} - i\varepsilon_0 \omega \vec{E} = (\underline{\underline{\sigma}} - i\varepsilon_0 \omega)\vec{E} \equiv -i\underline{\underline{\varepsilon}} \vec{E} . \tag{120}$$

The latter transformation corresponds to considering the currents induced by a wave as displacement current, which is considering plasma as a polarizable medium. Using (112) we obtain for the electric field

$$\left( k^2 - \vec{k}\vec{k} - \frac{\omega^2}{c^2} \underline{\underline{\varepsilon}} \right) \cdot \vec{E} = 0 . \tag{121}$$

Here $\vec{k}\vec{k}$ is the dyadic product of the vector $\vec{k}$ with itself. We obtain the dispersion relation from this system of equations by setting the determinant to zero. Writing formally $\underline{\underline{\mu\mu}} \equiv \vec{k}\vec{k}c^2/\omega^2$, we can formulate this as

$$\left| \mu^2 \underline{\underline{1}} - \underline{\underline{\mu\mu}} - \frac{\omega^2}{c^2} \underline{\underline{\varepsilon}} \right| = 0 . \tag{122}$$

We now have the problem of determining the dielectric tensor $\underline{\underline{\varepsilon}}$. For this purpose, we must determine $\underline{\underline{\sigma}}$. This is beyond the scope of this chapter, and we will give here only the (general) formula. For details see, for example, Refs. [10] and [11]. It is general use to abbreviate $\underline{\underline{\varepsilon}}$ in the form

$$\underline{\underline{\varepsilon}} \equiv \begin{Bmatrix} S & iD & 0 \\ -iD & S & 0 \\ 0 & 0 & P \end{Bmatrix}, \tag{123}$$

with the abbreviations

$$S \equiv 1 - \sum_k \frac{\omega_{pk}^2}{\omega^2} \frac{1 - i\nu_k/\omega}{(1 - i\nu_k/\omega)^2 - \omega_{Bk}^2/\omega^2} ,$$

$$D \equiv -\sum_k \frac{\omega_{pk}^2}{\omega^2} \frac{\omega_{Bk}}{\omega} \frac{1 - i\nu_k/\omega}{(1 - i\nu_k/\omega)^2 - \omega_{Bk}^2/\omega^2} , \tag{124}$$

$$P \equiv \sum_k \frac{\omega_{pk}^2}{\omega^2} \frac{1}{(1 - i\nu_k/\omega)} .$$

Here $k$ indicates the different kinds of charged particle (electrons and different kinds of ions), and $\omega_{pk}$ and $\omega_{Bk}$ are the plasma and gyro-frequencies respectively of species $k$.

To calculate the determinant, it is most convenient to introduce the coordinate system sketched in Fig. 16. In this system we have $\vec{B}_0 = (0.0, B_0)$ and $\vec{k} = (k\sin\theta, 0, k\cos\theta)$.

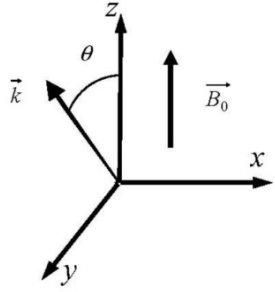

**Fig. 16:** Coordinate system used for Eq. (125)

With this notation our dispersion relation reads

$$\begin{vmatrix} \mu^2\cos^2\theta - S & -iD & -\mu^2\cos\theta\sin\theta \\ +iD & \mu^2 - S & 0 \\ -\mu^2\cos\theta\sin\theta & 0 & -\mu^2\sin^2\theta - P \end{vmatrix} = 0. \qquad (125)$$

When explicitly calculating Eq. (125), the $\mu^6$ terms vanish and we obtain a bi-quadratic expression for the refractive index,

$$A\mu^4 - B\mu^2 + C = 0, \qquad (126)$$

with

$$\begin{aligned} A &= S\sin^2\theta + P\cos^2\theta, \\ B &= RL\sin^2\theta + SP(1+\cos^2\theta), \quad R \equiv S+D,\ L \equiv S-D, \\ C &= PRL. \end{aligned} \qquad (127)$$

It has two solutions corresponding to *fast* and *slow* waves:

$$\mu_{f,s}^2 = \frac{B}{2A}\left(1 \pm \sqrt{1 - 4AC/B^2}\right). \qquad (128)$$

It is also possible to solve for $\theta$, yielding an expression known as the *Appleton–Lassen equation*

$$\tan\theta = \frac{-P(\mu^2 - R)(\mu^2 - L)}{(S\mu^2 - RL)(\mu^2 - P)}. \qquad (129)$$

This equation is most convenient for discussing wave propagation in the 'principal' directions $\theta = 0$ and $\theta = \pi/2$. Here $\theta = 0$ means propagation along $\vec{B}_0$, that is, $\vec{k} \parallel \vec{B}_0$. In this case also $\tan^2\theta = 0$, thus either

- $\mu_r \equiv \sqrt{R}$ (right circularly polarized waves) or
- $\mu_l \equiv \sqrt{L}$ (left circularly polarized waves).

For these waves we find in the approximation of high frequencies, $\omega \gg \omega_{Bi}$, a cut-off frequency with $\mu_{r,l} = 0$ for

$$\frac{\omega_p^2}{\omega^2} = 1 \pm \frac{\omega_{Be}}{\omega}. \tag{130}$$

Right circularly polarized waves have a resonance $\mu_r \to \infty$ at the electron cyclotron frequency $\omega_{Be}$, the 'electron cyclotron resonance', where the electric field vector rotates synchronously with the gyrating electrons. Left circularly polarized waves have no resonance at high frequencies. Instead, they become resonant with the gyrating ions at $\omega_{Bi}$.

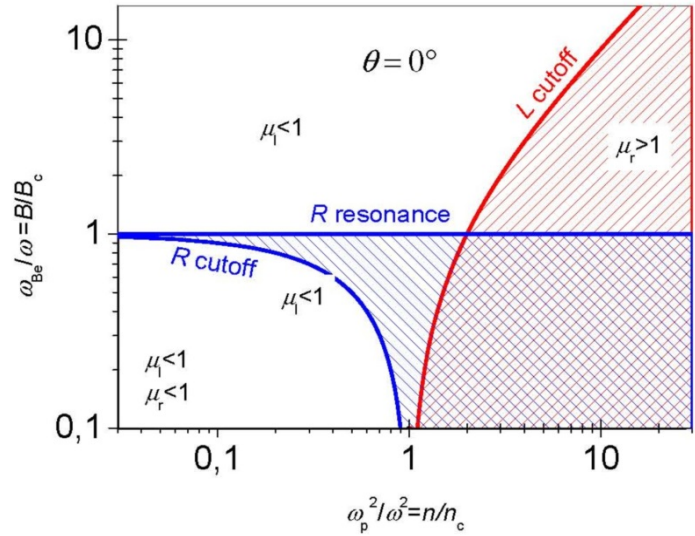

**Fig. 17:** Location curves for resonances and cut-offs at $\theta = 0°$ in a double-log $\omega_{Be}$ versus $\omega_p^2/\omega^2$ diagram (this is in principle a so-called CMA diagram).

For the discussion of wave propagation, one usually visualizes resonances and cut-offs by their localization curves in a diagram, where the abscissa displays $\omega_p^2/\omega^2 \equiv n/n_c$ and the ordinate $\omega_{Be}/\omega \equiv B/B_c$. Here $n_c$ and $B_c$ are called critical density and critical magnetic induction, respectively. These are, respectively, the density for which a given frequency is equal to the plasma frequency, and an induction for which a given frequency is equal to the electron gyro-frequency. Thus this diagram can be read either, for a given frequency, as a diagram of magnetic induction versus plasmas density or, for a given plasma, as a frequency diagram (or as a mixture of both). In this diagram Eq. (130) describes two straight lines, the same holding true for the resonance condition $\omega = \omega_{Be}$. This is shown in Fig. 17, where shaded regions are regions of no wave propagation along the magnetic field lines.

The different patterns indicate stop bands for right and left circularly polarized waves. The sharp resonances and cut-offs appear only when collisions are absent. This figure represents the basic structure of the so-called CMA diagram in the approximation $\omega_{Bi} = 0$.

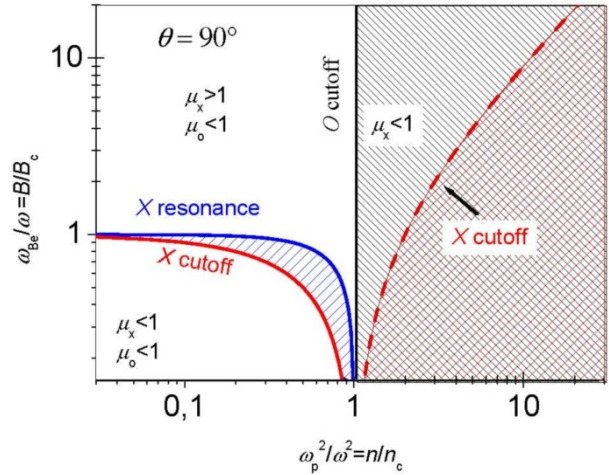

**Fig. 18:** Simplified CMA diagram for the case $\theta = 90°$ (see also Fig. 17)

In the case of waves propagating across the magnetic field, we have $\theta = 90°$, which implies $\tan\theta \to \infty$. This requires the denominator in Eq. (129) to become zero. For this case again we obtain two solutions: the waves with $\mu_o \equiv \sqrt{P}$ are called *ordinary waves*, while those with $\mu_x \equiv \sqrt{RL/S}$ are called *extraordinary waves*. The name *ordinary* is chosen to express that the refractive index for these waves does not depend on the external magnetic field. Ordinary waves have no resonance, only a cut-off at $\omega = \omega_p$ like waves in magnetic field-free plasma. The cut-offs for the extraordinary waves are the same as for the left and right circularly polarized waves – the localization curves are the same. They are given by $R = 0$ and $L = 0$. The resonance condition for the extraordinary wave is $S = 0$. As resonance frequency we find in the approximation of high frequencies

$$\frac{\omega_{Be}}{\omega} = \sqrt{1 - \frac{\omega_p^2}{\omega^2}} \ . \tag{131}$$

The respective curves are shown in Fig. 18. There is no wave propagation in the shaded regions. The pattern is used to distinguish between ordinary (o) and extraordinary (x) waves. While ordinary waves do not propagate for any frequency below $\omega_p$, extraordinary waves may propagate at lower frequencies if the magnetic field is sufficiently high.

To demonstrate the case of oblique direction of propagation, we show in Fig. 19 the case $\theta = 70°$. Here the designations X, O, R, L are extrapolations from the principal directions. Strictly speaking, these designations are defined only for the principal directions.

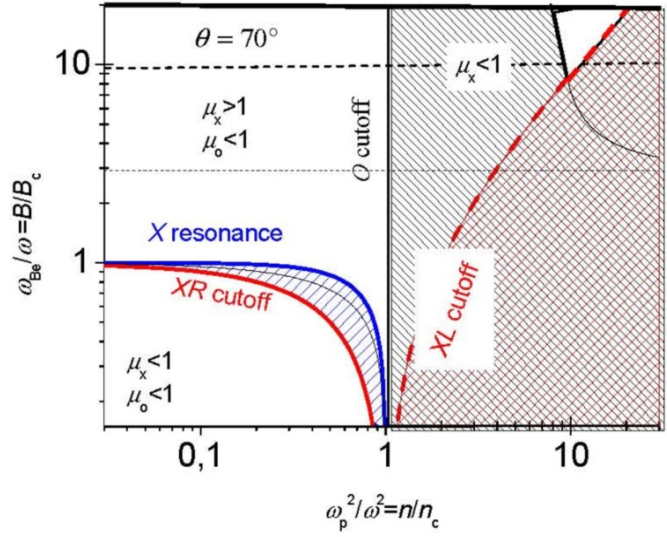

**Fig. 19:** Simplified CMA diagram for waves propagating at $\theta = 70°$ as an example for oblique propagation (see also Fig. 18).

What we have discussed so far were simplified versions of the so-called *Clemmow–Mullaly–Allis* (*CMA*) *diagram*, which helps in classifying plasma waves. In Fig. 20 we show a more complete version of this diagram. Here the ordinate represents $\omega_{Be}\omega_{Bi}/\omega^2$ instead of $\omega_{Be}/\omega$ to involve the ion effects for hydrogen ions, for instance the resonance of ion cyclotron waves (L-resonance). The curves shown are those for resonances and cut-offs at the principal directions of wave propagation (called *principal resonances* and *principal cut-offs* in the literature). These curves divide the diagram into 13 regions, in each of which exist topological distinct wave-normal surfaces (polar plots of the phase velocity $\overrightarrow{v_{\text{phase}}} = \omega \vec{k}/k^2$ versus the angle of propagation $\theta$). These polar diagrams are shown, but not to identical scales. Instead, the dotted circles represent *c*, the vacuum phase velocity of light, as an aid for orientation. Lemniscate shaped wave-normal surfaces indicate resonances with $v_{\text{phase}} = 0$. In any case these resonances occur in a distinct specific direction. The presentation is organized for the case that the *B* field vector points vertically upwards.

The CMA diagram is very helpful, for example, in discussing the accessibility of a plasma by a certain wave. Imagine the problem of launching a wave from an antenna outside plasma. 'Outside' means that the plasma density *n* is almost zero. If, also, the magnetic field is zero and rises inside the plasma, the conditions at the location of the antenna correspond to the left lower corner of the diagram. This means that the wave will propagate into a stop band limited by a cut-off and cannot penetrate further into the plasma. A possible solution is to install at the antenna a magnetic field with $B > B_c$, thus creating conditions where waves can propagate.

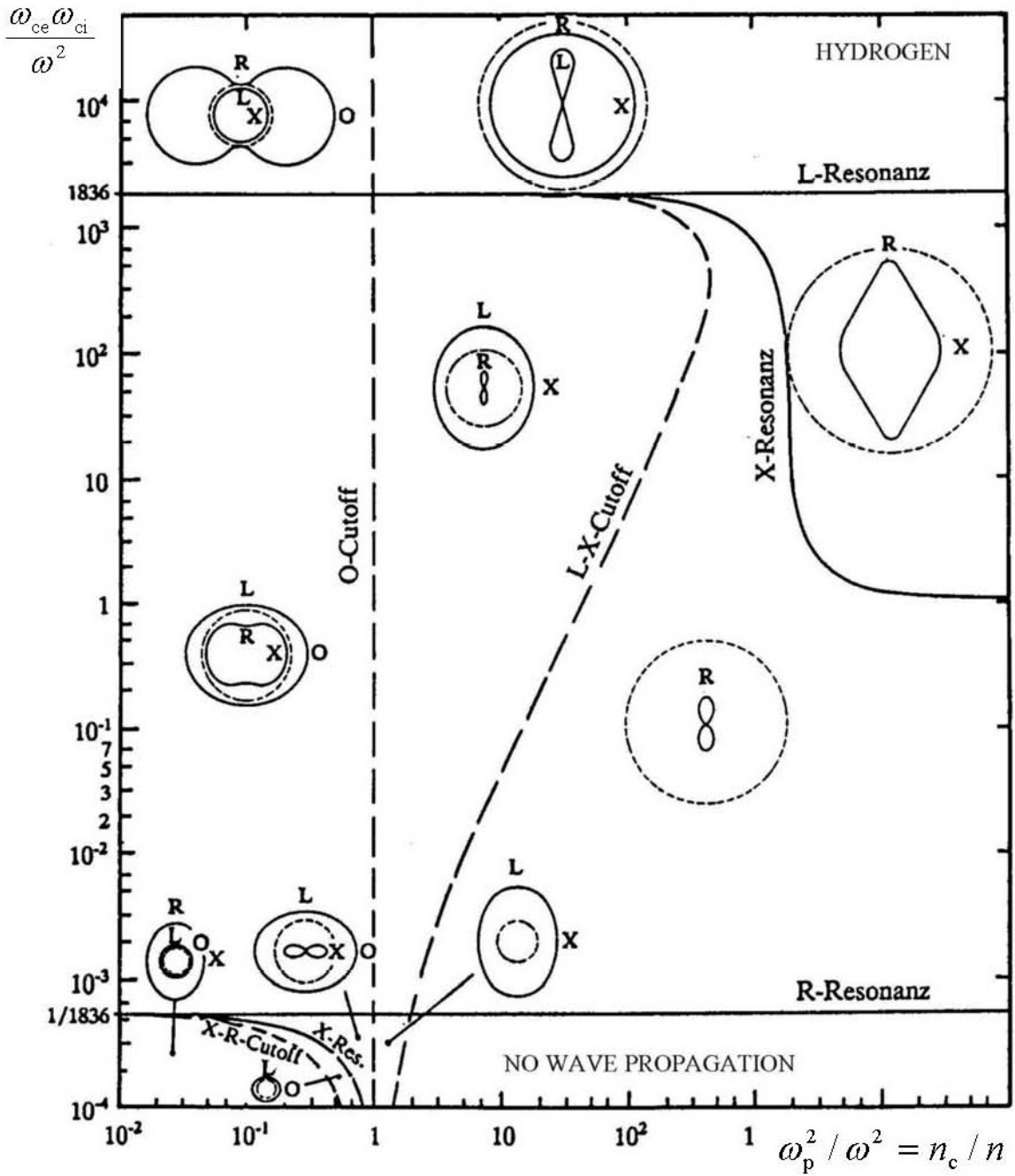

**Fig. 20:** Complete CMA diagram including ion effects for H$^+$ ions, such as, for example, the L-resonance for left circularly polarized ion cyclotron waves, and further phase velocity polar diagrams. (From Ref. [14], © G. Jansen, Wangen, Germany, reprinted with permission.)

## 5   Concluding remarks

Here ends our short excursion into the world of plasma. It was necessarily short and in no way complete. To learn more, one needs to study the literature. There are a number of introductory plasma physics books on the market. In my opinion Ref. [9] is one of the best, at least for a reader new to the field, because it is easy to read and gives a very good overview. However, I am not familiar with the very recently published books, and for those I cannot give a recommendation. In general, these books emphasize the problems of fully ionized plasma as studied in fusion research and astrophysics. Plasma waves are discussed in much more detail in the book of Stix [10]. Our presentation further owes much to the treatment of wave phenomena in Refs. [10]–[13]. Jansen's book [14] gives many helpful diagrams on wave behaviour in plasma. The special problems of ion source plasma are discussed in a short review in Ref. [15]. Many ion source discharges have much in common with discharges for material processing. The book of Lieberman and Lichtenberg [16] gives a very concise and well-written treatment of the problems of those plasmas and may be found extremely useful when studying the physics of ion source discharges. Last, but not least, the book of Geller [8], the 'father of ECRIS' (electron cyclotron resonance ion source), is a very personal report of the author's involvement and struggle in developing the ECRIS principle and understanding the physics of ECRIS plasma.